\documentclass[twocolumn]{aastex631}
\usepackage{ragged2e}

\usepackage{siunitx}
\usepackage{booktabs}
\usepackage{graphicx}

\usepackage{xspace}
\usepackage{apjfonts}
\usepackage{float}
\usepackage{hyperref}
\definecolor{LightCyan}{rgb}{0.88,1,1}
\usepackage{color, colortbl}
\newcommand{\thisstar}{TIC 88785435} 
\newcommand{\thisstarb}{TIC 88785435\,b}
\newcommand{\radb}{$R_b = 5.03_{-0.20}^{+0.21}\,\mathrm{R_\oplus}$}

\newcommand{\shortperb}{$P_b = 10.51$ days}

\newcommand\mysim{\mathord{\sim}}

\newcommand\vsini{\ifmmode{v\sin{i_\star}}\else $v\sin{i_\star}$\fi}

\newcommand\sini{\ifmmode{\sin{i_\star}}\else $\sin{i_\star}$\fi}

\newcommand{\tess}{{\it TESS}}

\newcommand{\cfa}{Center for Astrophysics \textbar{} Harvard \& Smithsonian, 60 Garden Street, Cambridge, MA 02138, USA}

\newcommand{\usq}{Centre for Astrophysics, University of Southern Queensland, West Street, Toowoomba, QLD 4350, Australia}
\newcommand{\ames}{NASA Ames Research Center, Moffett Field, CA, 94035, USA}

\newcommand{\princeton}{Department of Astrophysical Sciences, Princeton University, 4 Ivy Lane, Princeton, NJ 08544, USA}

\newcommand{\dartmouth}{Department of Physics and Astronomy, Dartmouth College, Hanover, NH 03755, USA}

\newcommand{\nexsci}{Caltech/IPAC -- NASA Exoplanet Science Institute 1200 E. California Ave, Pasadena, CA 91125, USA}

\newcommand{\jpl}{Jet Propulsion Laboratory, California Institute of Technology, Pasadena, CA 91109 USA}
\usepackage{amsmath}

\shorttitle{A newborn super-Neptune in Sco-Cen}
\shortauthors{Vach et al.}

\graphicspath{{./}{figures/}}
\begin{document}
% \linenumbers
\title{A 16 Myr super-Neptune in Upper-Centaurus Lupus and a preliminary survey of transiting planets in Sco-Cen with TESS}

%%%%%%%%%%%%%%%%%%%%%%%%
% Direct contributions
%%%%%%%%%%%%%%%%%%%%%%%%

\author[0000-0001-9158-9276]{Sydney Vach}\altaffiliation{Corresponding author: \href{mailto:sydney.vach@unisq.edu.au}{sydney.vach@unisq.edu.au} }%
\affiliation{\usq}

\author[0000-0002-4891-3517]{George Zhou}\affiliation{\usq} %
% \affiliation{\usq}

\author[0000-0003-3654-1602]{Andrew W. Mann}%
\affiliation{Department of Physics and Astronomy, The University of North Carolina at Chapel Hill, Chapel Hill, NC 27599, USA}

\author[0000-0002-8399-472X]{Madyson G. Barber}%
\altaffiliation{NSF Graduate Research Fellow}
\affiliation{Department of Physics and Astronomy, The University of North Carolina at Chapel Hill, Chapel Hill, NC 27599, USA}

\author[0000-0002-0692-7822]{Tyler R. Fairnington}\affiliation{\usq}%

\author[0000-0003-0918-7484]{Chelsea X. Huang}\affiliation{\usq}

\author[0000-0001-7615-6798]{James G. Rogers}
\affiliation{Institute of Astronomy, University of Cambridge, Madingley Road, Cambridge CB3 0HA, UK}%

\author[0000-0002-0514-5538]{Luke G. Bouma} %
\affiliation{Observatories of the Carnegie Institution for Science, Pasadena, CA 91101, USA}
% \altaffiliation{51 Pegasi b Fellow}
% 

\author[0009-0003-3841-5383]{Joachim Kr{\"u}ger}
\affiliation{\usq}%

\author{Duncan Wright}%
\affiliation{\usq}

\author[0009-0005-2363-9274]{Annabelle E. Niblett}%
\affiliation{\dartmouth}

\author[0009-0009-4540-4803]{Jack M. Nelson}
\affiliation{\dartmouth}%

\author[0000-0002-8964-8377]{Samuel N. Quinn} 
\affiliation{\cfa}

\author[0000-0001-9911-7388]{David W. Latham}%
\affiliation{\cfa}

\author[0000-0001-6637-5401]{Allyson Bieryla}
\affiliation{\cfa}
\affiliation{\usq}

\author[0000-0001-6588-9574]{Karen A. Collins}%
\affiliation{\cfa}

\author[0000-0001-9269-8060]{Michelle Kunimoto}%
\affiliation{Department of Physics and Astronomy, University of British Columbia, 6224 Agricultural Road, Vancouver, BC V6T 1Z1, Canada}

\author[0000-0001-8621-6731]{Cristilyn N.\ Watkins}%
\affiliation{\cfa}

\author[0000-0001-8227-1020]{Richard P. Schwarz}%
\affiliation{\cfa}

\author[0000-0003-2781-3207]{Kevin I. Collins}%
\affiliation{George Mason University, 4400 University Drive, Fairfax, VA, 22030 USA}

\author[0000-0003-3904-6754]{Ramotholo Sefako}%
\affiliation{South African Astronomical Observatory, P.O. Box 9, Observatory, Cape Town 7935, South Africa}

\author[0000-0003-1728-0304]{Keith Horne}%
\affiliation{SUPA Physics and Astronomy, University of St\,Andrews, Fife, KY16 9SS Scotland, UK}

\author[0000-0002-2532-2853]{Steve B. Howell}%
\affiliation{\ames}

\author[0000-0002-2361-5812]{Catherine~A.~Clark}%
\affiliation{\jpl}
\affiliation{\nexsci}

\author[0000-0001-7746-5795]{Colin Littlefield}%
\affiliation{Bay Area Environmental Research Institute, Moffett Field, CA 94035, USA}

\author[0000-0002-8035-4778]{Jessie L. Christiansen}%
\affiliation{\nexsci}

\author[0000-0002-2482-0180]{Zahra Essack}%
\affiliation{Department of Physics and Astronomy, The University of New Mexico, 210 Yale Blvd NE, Albuquerque, NM 87106, USA}

\author[0000-0002-4265-047X]{Joshua N. Winn}%
\affiliation{\princeton}

\begin{abstract}
Measuring the properties of planets younger than about $50$ Myr helps to test different planetary formation and evolution models. NASA's Transiting Exoplanet Survey Satellite (\tess) has observed nearly the entire sky, including a wide range of star-forming regions and young stellar clusters, expanding our census of the newborn planet population. In this work, we present the discovery of the \thisstar\ planetary system located in the Upper-Centaurus Lupus (UCL) region of the Scorpius-Centaurus OB association (Sco-Cen) and a preliminary survey of the planet population within Sco-Cen. \thisstar\ is a pre-main sequence, K7V dwarf ($M_\star = 0.72\,M_\odot$, $R_\star = 0.91\,R_\odot$, $T_\mathrm{eff} = 3998\,$K, $V$ = 11.7 mag) located within the bounds of UCL. We investigate the distribution of rotation periods measured from the TESS long-cadence data and the H$\alpha$ and Li abundances from the spectra of \thisstar. \tess\ long-candence data reveal that \thisstar\ hosts a transiting super-Neptune ($R_b = 5.03\,R_\oplus$, $P =10.51 $ days), \thisstarb. Ground-based follow-up validates the planetary nature of \thisstarb. Using the \tess\ data, we perform a preliminary survey to investigate how \thisstarb\ compares to the population of newly born planets located within Sco-Cen.
\end{abstract} 

\keywords{Exoplanets (498); Mini Neptunes (1063); Transit photometry (1709); Young star clusters (1833); Exoplanet astronomy (486)}
\section{Introduction}

%The exoplanet population 

Stellar clusters and associations have proven to be excellent hunting grounds for young transiting planets. Planets orbiting stars still located within their birth population enable age measurements with a higher degree of precision than field stars. In turn, the precise ages of these exoplanetary systems allow for a more accurate comparison between our theories of planet formation and evolution and the properties of the observed young exoplanet population. Specifically, the first tens of Myr post-formation are predicted to feature drastic processes of planetary evolution-- e.g., thermal contraction \citep[][]{Lopez:2012}, atmospheric mass loss via photoevaporative escape \citep[][]{owen:2013,Owen:2017}, and orbital migration \citep[e.g.,][]{Ida:2004, Lee:2014, Lambrechts:2019}.

NASA's \textit{K2} \citep[][]{k2} mission observed various well-characterized stellar associations along the ecliptic plane, e.g., Upper-Sco, Pleiades, Praesepe, Hyades. These young transiting planet systems (e.g., K2-33--Upper Sco; \citet{David:2016, Mann2017}{}{}, V1298 Tau--Taurus Auriga; \cite{David:2019}{}{}) provided an initial glimpse into the differences between the young and mature exoplanet populations \citep[][]{Rizzuto:2017, Christiansen:2023}{}{}. Now, NASA's Transiting Exoplanet Survey Satellite \citep[\tess;][]{TESS}{}{} has observed nearly the entire sky, providing near-complete coverage of bright members of well-characterized associations. \tess\ has greatly expanded the known population of young exoplanets in clusters and associations (e.g., DS Tuc A \citep[Tuc-Hor;][]{Newton2019}{}{}, AU Mic \citep[$\beta$ Pic][]{Plavchan2020}{}{}, HIP 67522 \citep[UCL;][]{Barber:2024}{}{}, HD 109833 \citep[MELANGE-4;][]{Wood:2023}{}{}, IRAS 04125+2902 b \citep[Taurus Molecular Cloud;][]{Barber:IRAS} ). 

The youngest \tess\ planet population further illuminated the differences between young and mature exoplanets. \tess\ revealed a distribution of young planets with larger radii, hinting at a gas-rich formation scenario \citep[][]{Vach:2024,Fernandes:2022, Karalis:2025}{}{}. JWST and HST observations of the young Jovian-sized planets, HIP 67522 b \citep[][]{Thao:2024} and V1298 Tau b and c \citep[][]{Barat:2024a, Barat:2024b}, further support this scenario, as all three mass measurements are consistent with super-Earths and mini-Neptunes rather than young Jovian-mass planets. The recent dynamical mass determination of the slightly older (35 Myr), Saturn mass planet TOI-837 b \citep[][]{Bouma:2020, Barragan:2024, Damasso:2024} illustrates the need for mass characterization of the young planet population, as well as further young planet searches targeting young comoving populations and star forming regions.

As the largest nearby association with recent star formation, Sco-Cen provides an ideal set of targets to search for young transiting planets with well-characterized ages, which all evolved from similar formation environments. Here, we present the discovery and characterization of \thisstarb, a 16 Myr super-Neptune orbiting the established pre-main sequence member of the Sco-Cen. We also present our initial survey of Sco-Cen with \tess. In Section~\ref{sec:obs}, we present an overview of our planet search procedure and our observations of \thisstar. We spectroscopically characterize \thisstar, and analyze its membership within the Sco-Cen population in Section~\ref{sec:star}. The global modeling procedure is presented in Section~\ref{sec:global_model}, and our false-positive analysis is presented in Section ~\ref{sec:fp}. Section~\ref{sec:survey} presents the initial results from our magnitude-limited Sco-Cen planet survey with \tess. Our discussion and conclusions are in Section~\ref{sec:discussion}.

% \[\]
% Train of thought:\\
% - stellar clusters are great we get ages\\
% - if we have ages, we can compare to what theory predicts the population should look like\\
% - k2 found a few young planets, but really most were a bit older, but then tess came in and stole the show\\
% - In the era of JWST, we can combine both population statistics and direct observations of young exoplanet atmosphere to test our theories of planet formation and evolution
% - Here, we present a new youngin that is adding to the young population that can be used to test theories

\section{Observations}\label{sec:obs}

\subsection{\emph{TESS} Photometry}
NASA's \tess\ mission is an all-sky survey dedicated to hunting for transiting exoplanets around nearby, bright stars. \tess\ stores data for all sources within its entire field of view in Full Frame Images (FFIs). Within \tess's primary mission, FFIs were collected at 30-minute cadences. FFIs were then sampled at 10-minute and then 200-second cadences in the first and second extended missions. 

\tess\ observed \thisstar\ across Sectors 11, 38, and 65 in the FFIs. These data were processed by the Quick Look Pipeline \citep[][]{qlp2020a,qlp2020b}{}{} and made available via the Mikulski Archive for
Space Telescopes (MAST). \thisstar\ was originally identified as a planet candidate host in \citet{Vach:2024}, which surveyed young associations for transiting planets. Briefly, \citet{Vach:2024} performed a flare rejection via an iterative sigma clipping on the \tess\ light curves. Light curves were then detrended using a spline following \citet{Vanderburg:2019} and searched for transit signals with a Box-least squares \citep[BLS; ][]{kovacs:2002} search. A signal-to-pink noise ratio $\geq8$ was required to trigger a threshold-crossing event (TCE).

% All TCEs were then visually examined, and subjected to a centroid analysis, even-odd transit depth analysis, and pixel-by-pixel analysis to determine if the signal was on target, consistent with a planetary signal, and rule out any obvious astrophysical false-positive. TCEs that passed the vetting procedures were then checked for multiplicity. 

\begin{figure*}
    \centering
    \includegraphics[width=\linewidth]{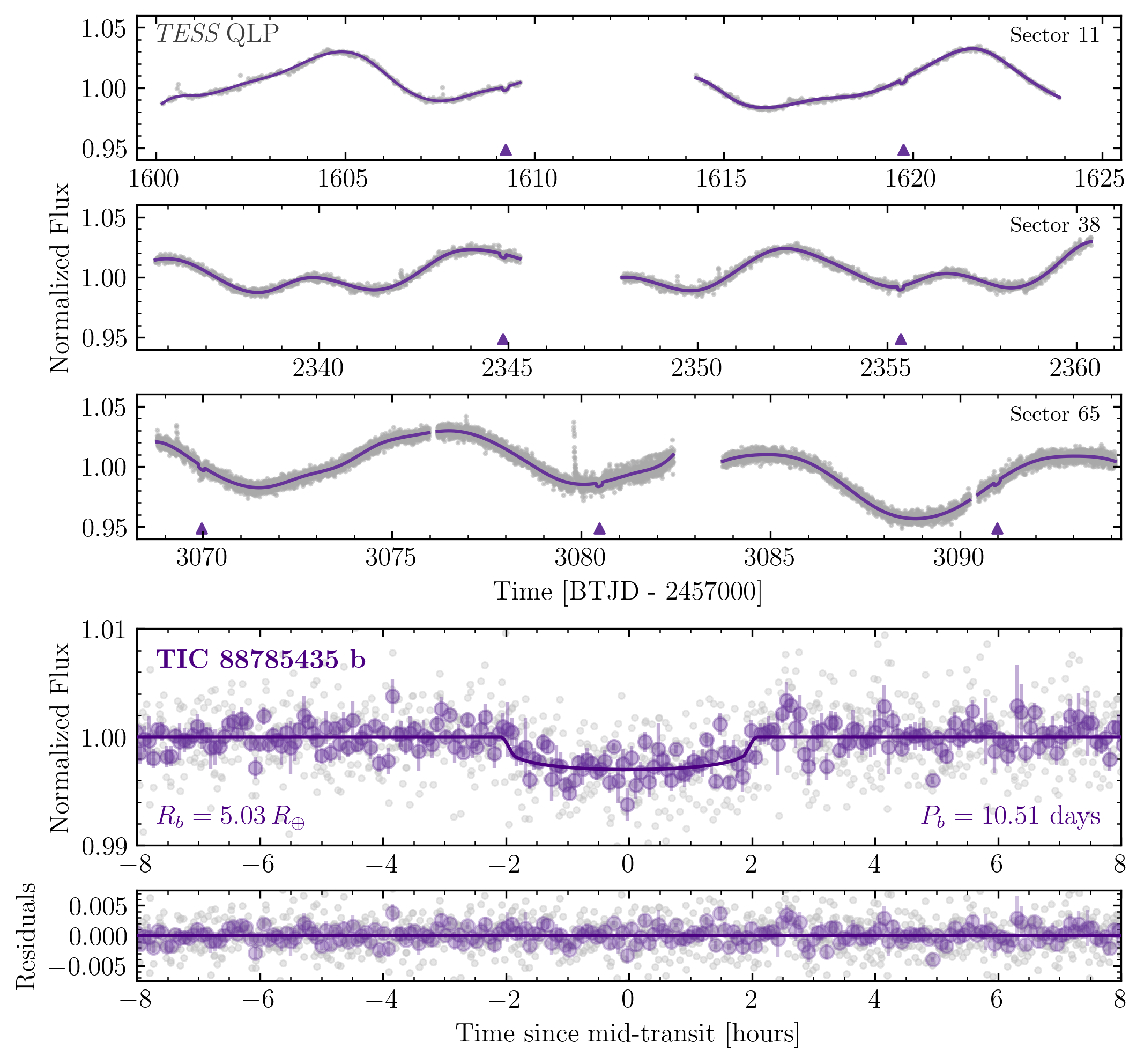}
    \caption{\textit{Top three panels:} \tess\ FFI light curves (grey) for \thisstar\ across Sectors 11, 38, and 65. We model the stellar activity using a spline fit and the transits are modeled with \texttt{batman} using our best-fit parameters from our global model (purple). The transits of \thisstarb\ are marked by purple triangles. \textit{Bottom two panels:} Phase-folded \tess\ transit and best-fit model of the super-Neptune, \thisstarb, and the model residuals. }
    \label{fig:tess}
\end{figure*}

A 10.51 day periodic event in the FFI light curves of \thisstar\ triggered a TCE with a signal-to-pink noise ratio of 9.47, passing all vetting procedures. \thisstarb\ was released to the community as a CTOI on 2024-01-19 UTC. In Figure \ref{fig:tess}, we present the full \tess\ light curve and the phase-folded transit of \thisstarb. The \tess\ light curves exhibit significant rotational modulation with a period of $\mysim 8\,$ days and an amplitude of $\mysim3\%$.  %In Section~\ref{sec:global_model}, we describe our global model used to determine the best-fit parameters and explore their posteriors. 

\subsection{Ground-based Photometry}

We observed four transits of \thisstarb\ through the \tess\ Follow-up Observing Program \citep[TFOP;][]{collins:2019}{}{}\footnote{https://tess.mit.edu/followup} with the 1-meter at Las Cumbres Observatory \citep[LCO; ][]{Brown:2013}{}{}{} telescope located at the Cerro Tololo Inter-American Observatory (LCO-CTIO), Cerro Tololo, Chile. LCO-CTIO is equipped with a $1024\times1024$ SINISTRO camera. The SINISTRO camera is a $4$ which observes a 13’ x 13’ field of view with a pixel scale of 0.778.'' We present a summary of our LCO-CTIO observations in Table~\ref{tab:lco}.

LCO-CTIO observed two egresses of \thisstarb\ on 2024-02-13 UTC and 2024-03-05 UTC in the $i'$ band. These observations were scheduled using a customized version of the \texttt{Tapir} package \citep[][]{Jensen:2013}{}{}, \texttt{TESS Transit Finder}. A full transit was observed in the $i'$ band on 2024-03-26 UTC. Another full transit was observed on 2024-04-16 UTC in the $g'$ band. 

\begin{figure}
    % \centering
    \includegraphics[width=0.93\linewidth]{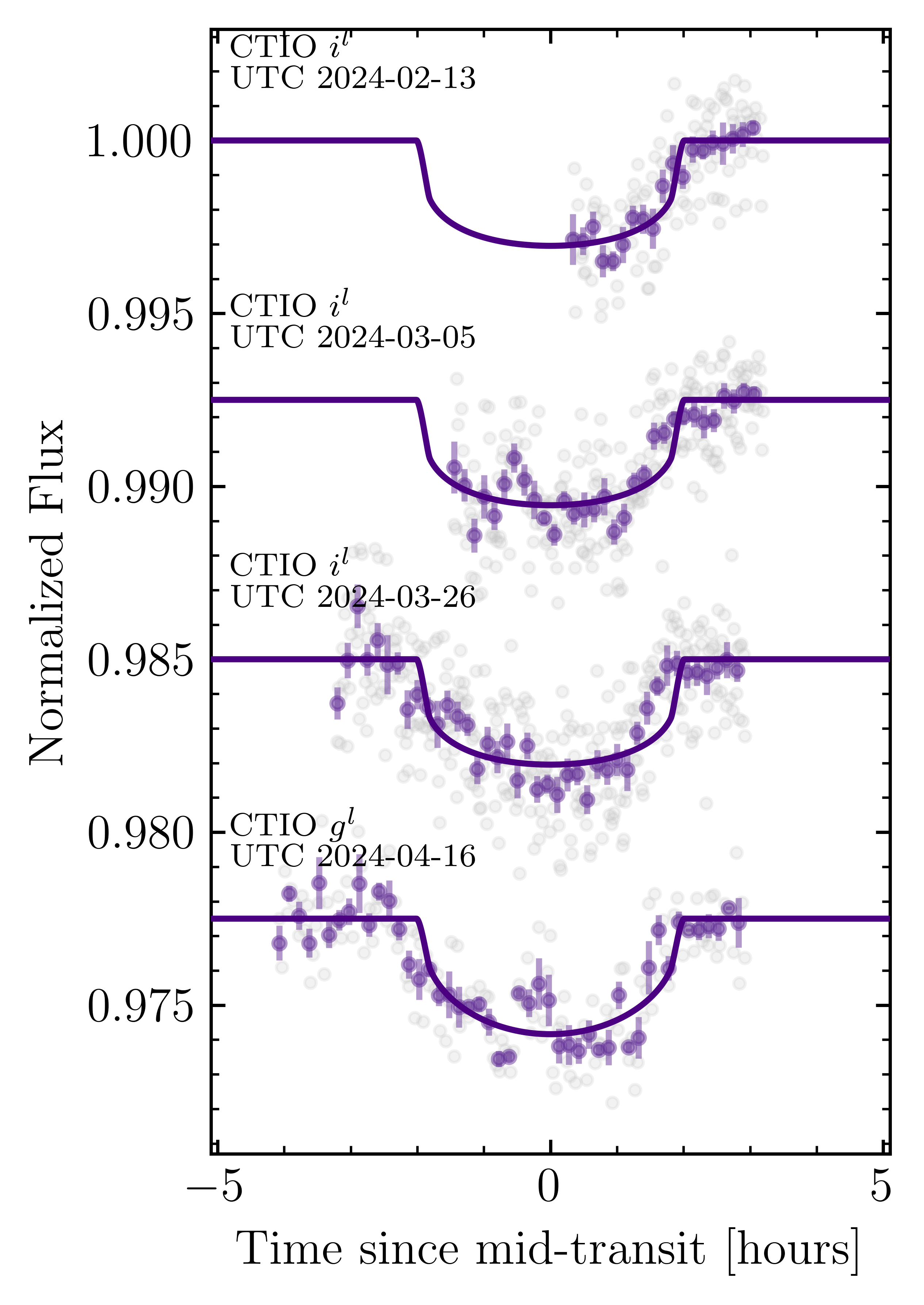}
    \caption{Phase-folded ground-based transit observations of \thisstarb\ (grey, binned in purple). Overlaid are our best-fit transit models via \texttt{batman} for each observation.}
    \label{fig:lco}
\end{figure}

The LCO-CTIO images were calibrated using LCO \texttt{BANZAI} \citep[][]{McCully:2018}{}{}, and the light curves were extracted using \texttt{AstroImageJ} \citep[][]{Collins:2017}{}{}. We included all LCO raw photometry and performed a simultaneous detrending in our global modeling procedure. The detrended LCO transits are presented in Figure~\ref{fig:lco}. The LCO light curves confirm the \tess\ transits with similar depth and duration in both the $g'$ and $r'$ bands. All LCO observations were used in our global model (see Section~\ref{sec:global_model}). 

% \begin{table}
% 	% \centering
% 	\caption{Ground-based photometry of \thisstar.}
% 	\label{tab:lco}
% 	\begin{tabular}{llccl} % four columns, alignment for each
% 		\hline
% 		\,\,\,Date (UTC) & Instrument & Duration & Aperture &Filter\\
% 		\hline
% 		\textbf{TIC 434398831 b}\\
%             \,\,\,2023-09-28 &LCO-TEID 1.0\,m  & egress & 3.1'' & \textit{i'} \\
% 		\,\,\,2023-11-25 &LCO-SAAO 1.0\,m & ingress & &\textit{i'}\\
% 		\,\,\,2023-11-25  &LCO-TEID 1.0\,m$^*$&  full & 5.1''& \textit{i'}\\
%             \,\,\,2023-12-18  &LCO-CTIO 1.0\,m&  full & & \textit{i'}\\
%             \,\,\,2023-12-18  &LCO-TEID 1.0\,m&  full & &\textit{i'}\\
%             \,\,\,2024-02-11  &MuSCAT3 2.0\,m&  full & &\textit{i', g', r',} $z_s$\\
%             \,\,\,2024-02-14  &Whitin 0.7\,m&  full & &\textit{R}\\
%             \textbf{TIC 434398831 c}\\
%             \,\,\,2023-09-03 &LCO-McD 1.0\,m&  egress & 4.3'' & \textit{i'} \\
%             \,\,\,2023-09-28 &LCO-TEID 1.0\,m&  ingress & 3.1''& \textit{i'} \\
%             \,\,\,2023-12-17 &LCO-SAAO 1.0\,m&  full & &\textit{i'} \\
%             \,\,\,2023-12-30 &MuSCAT3 2.0\,m&  full & &\textit{i', g', r',} $z_s$ \\
% 		\hline
% 	\end{tabular}
%  $^*$\textit{Not included in global model due to poor weather conditions.}
% \end{table}

\begin{table}
	% \centering
	\caption{Ground-based photometry of \thisstar.}
	\label{tab:lco}
	\begin{tabular}{lcccc} % four columns, alignment for each
        \hline\hline
		\,\,\,Instrument & Date (UTC) & Duration & Aperture &Filter\\
		\hline
            \,\,\,LCO-CTIO 1.0\,m&2024-02-13  & Egress & 5.1''& \textit{i'}\\
            \,\,\,LCO-CTIO 1.0\,m&2024-03-05  & Egress & 4.7''& \textit{i'}\\
            \,\,\,LCO-CTIO 1.0\,m&2024-03-26  & Full & 3.9''& \textit{i'}\\
            \,\,\,LCO-CTIO 1.0\,m&2024-04-16  & Full & 5.5''& \textit{g'}\\
            		\hline

	\end{tabular}

\end{table}\label{tab:lco}

\subsection{Spectroscopic follow-up}\label{sec:spec}

We obtained seven observations of \thisstar\ using the Veloce spectrograph on the 3.9-m Anglo Australian Telescope, located at Siding Spring Observatory, Australia \citep{veloce}. Veloce is an echelle spectrograph with a resolving power of $R\sim75,000$, fed via a fiber bundle 2.4" in diameter. We made use of observations from the red-optical arm of Veloce over the wavelength range of 5800-9300\,\AA{}, which has the highest instrument throughput. %Additionally, we obtained the spectra of 6 nearby stars, which are likely members of the same subpopulation of Sco-Cen (see Section\.\ref{sec:li_halpha}).

Observations were reduced as per J. Kr{\"u}ger et al. \textit{in prep}. Relative wavelength calibration was performed via bracketing Thorium-Uranium-Argon lamp exposures about each science exposure. Stellar line broadening profiles were derived via a least-squares deconvolution \citep{Donati1997} against a non-rotating synthetic template generated via the ATLAS9 atmosphere models \citep{Castelli:2004}. Radial velocities and line broadening velocities were derived from each exposure by modeling the line broadening profiles as per \citet{Gray:2005}. 

We find the velocities of \thisstar\ to be stable at the $\sim 0.6\,\mathrm{km\,s}^{-1}$ level, limited by the instrument stability with this existing reduction pipeline. The velocities are presented in Table~\ref{tab:rv}. The line profile shows velocity broadening at the limit of the spectral extraction induced broadening $(<10\,\mathrm{km\,s}^{-1})$, consistent with the slow rotation of the star, and no signs of double-lined features.

We obtained the spectra of 6 nearby stars, which have been identified as likely members of Sco-Cen with no previous spectroscopic observations. These stars are likely members of the same subpopulation as \thisstar. The observations were taken using the same observing strategy and reduced using the same methodology as described above. We expand upon their characterization in Section\,\ref{sec:li_halpha}.

%6 nearby stars that are likely members of the same group. the observations were taken in the same way as for the target, and the SNR was good
\begin{table}
    \caption{Radial Velocities\label{tab:rv}}
    \centering
    \begin{tabular}{cccr}
        \hline\hline
         BJD & RV [km\,s$^{-1}$] & RV Error [km\,s$^{-1}$] & Instrument \\
        \hline
         2460451.12442 & 0.1 & 2.0 & Veloce\\
         2460451.93020 & 0.84 & 0.64 & Veloce\\
         2460452.06695 & 1.31 & 0.71 & Veloce\\
         2460452.22589 & -0.5 & 1.7 & Veloce\\
         2460453.00587 & 1.33 & 0.51 & Veloce\\
         2460453.11426 & 0.55 & 0.51 & Veloce\\
         2460453.24952 & 0.05 & 0.34 & Veloce\\
         \hline
    \end{tabular}
\end{table}

\subsection{High-Resolution Imaging}\label{sec:hri}
\begin{figure}
    % \centering
    %\includegraphics[width=\linewidth]{hri.png}
    \includegraphics[width=\linewidth]{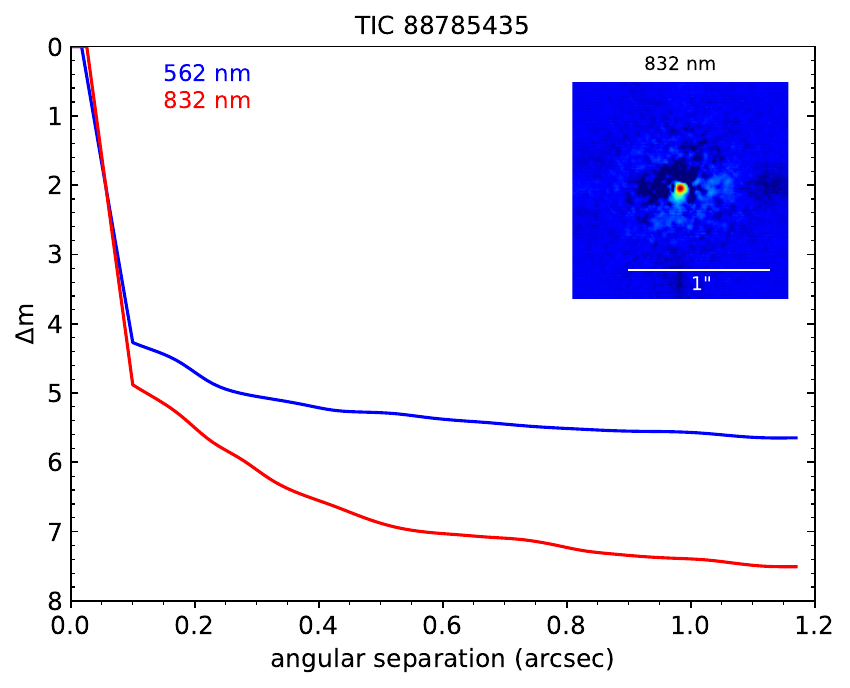}
    \caption{Gemini South 8\,m blue (562 nm) and red (832 nm) arm diffraction limited images and 5$\sigma$ contrast curves of \thisstar. No companions were found within detection limits.}
    \label{fig:hri}
\end{figure}

We obtained high-resolution images of \thisstar\ using the Zorro speckle imager on the 8-meter Gemini South telescope \citep[][]{Scott:2021}{}{} to search for visual companions. The Zorro speckle imager has a blue and red arm, centered at 562 nm and 832 nm respectively, with a field of view of $ 2.5''\times2.5''$ and a pixel scale of 0.01'' per pixel. The data were reduced following \citet{Howell:2011}. No secondary sources were detected around 
\thisstar.
We obtained contrast ratios of $\Delta m >5$ mag at 0.1'' and $\Delta m >7$ mag at 0.5'' in the red arm. We present our $5\sigma$ contrast curves in Figure~\ref{fig:hri}.

\section{Stellar Characterization}\label{sec:star}
\begin{deluxetable}{lcr}

\tablewidth{0pc}
\tabletypesize{\scriptsize}
\tablecaption{
        Properties of \thisstar% \hatstara{}
    \label{tab:star}
}
\tablehead{
    &
    &
    \\
    \multicolumn{1}{l}{~~~Paramter}          &
    \multicolumn{1}{c}{Value}            &
    \multicolumn{1}{r}{Source}            
}
\startdata  
\multicolumn{3}{l}{\textbf{Identifier}}\\ 
            ~~~TIC ID & TIC 88785435 & 1\\
             ~~~2MASS & J14570814-3052476 & 2\\
             ~~~APASS &  16447752 & 3\\
             ~~~Gaia DR3 & 6205812887538362624 & 4\\
             ~~~UCAC4& 296-080560 & 5\\
             ~~~WISE & J145708.13-305247.9 & 6\\
		\textbf{Astrometry}\\
             ~~~Right Ascension (RA) \dotfill &  14:57:08.13 & 4\\
             ~~~Declination (Dec) \dotfill & -30:52:48.1 & 4\\
             ~~~Parallax (mas) \dotfill & 8.191$\pm$0.015& 4\\
             \textbf{Proper Motion}\\
             ~~~RA Proper Motion (mas yr$^{-1}$) \dotfill & $-24.601\pm0.019$ & 4\\
             ~~~Dec Proper Motion (mas yr$^{-1}$) \dotfill &$-26.356\pm0.015$ & 4\\
             \textbf{Photometry}\\
             ~~~\tess\ (mag) \dotfill &$11.7279 \pm 0.0072$ & 1\\
             ~~~\emph{B} (mag) \dotfill & $14.659 \pm 0.059$& 3\\
             ~~~\emph{V} (mag) \dotfill &$13.259 \pm 0.114$ & 3\\
             ~~~\emph{J} (mag) \dotfill & $10.435 \pm 0.026$ & 2\\
             ~~~\emph{H} (mag) \dotfill & $9.764 \pm 0.027$ & 2\\
             ~~~\emph{K} (mag) \dotfill & $9.546 \pm 0.021$ & 2\\
             ~~~\emph{Gaia} (mag) \dotfill & $12.6048 \pm 0.0015$ & 4\\
             ~~~$\mathrm{\emph{Gaia}}_\mathrm{BP}$ (mag) \dotfill &13.5145$\pm 0.0054$& 4\\
             ~~~$\mathrm{\emph{Gaia}}_\mathrm{RP}$ (mag) \dotfill & 11.6590 $\pm 0.0034$ & 4\\
              ~~~WISE W1 (mag) \dotfill & $9.446 \pm 0.022$& 6\\
              ~~~WISE W2 (mag) \dotfill & $9.453 \pm 0.02$& 6\\
              ~~~WISE W3 (mag) \dotfill &$9.34 \pm 0.045$ & 6\\
              ~~~WISE W4 (mag) \dotfill & $>8.329$ & 6\\
              \textbf{Physical Properties}\\
              ~~~$M_\star$ (M$_\odot$) \dotfill & 0.724$\pm$0.017&7\\
              ~~~$R_\star$ (R$_\odot$) \dotfill & 0.911$\pm$0.038& 7\\
              ~~~$T_\mathrm{eff}$ (K) \dotfill & 3998$\pm$95 & 7\\
              ~~~Surface gravity $\log g_\star$ (cgs) \dotfill & 4.19$\pm$0.37& 7\\
              ~~~[m/H] \dotfill & -0.08$\pm$0.11 & 7\\
              %~~~$v\sin I_\star$ ($\mathrm{km\,s^{-1}}$) \dotfill & 10.07$\pm$4.14 & 7\\
              ~~~$v\sin i_\star$ ($\mathrm{km\,s^{-1}}$) \dotfill & $<10$ & 7\\
              ~~~Age (Myr) \dotfill & 16.0$\pm$1.6 & 7\\
              ~~~Distance (pc) \dotfill & 122.08$\pm$0.25 & 7\\
              \textbf{Limb darkening coefficients}\\
              ~~~$u_{1,\mathrm{\tess}}$ \dotfill  & 0.349$\pm$0.026& 8\\
              ~~~$u_{2,\mathrm{\tess}}$ \dotfill  & 0.256$\pm$0.014& 8\\
              ~~~$u_{1,\mathrm{\emph{i'}}}$ \dotfill  & 0.364$\pm$0.030& 8\\
              ~~~$u_{2,\mathrm{\emph{i'}}}$ \dotfill  & 0.238$\pm$0.020& 8\\
              ~~~$u_{1,\mathrm{\emph{g'}}}$ \dotfill  & 0.656$\pm$0.049& 8\\
              ~~~$u_{2,\mathrm{\emph{g'}}}$ \dotfill  & 0.139$\pm$0.045& 8\\
              \textbf{Activity indicators}\\
               ~~~$P_{\mathrm{rot}}$ (days) \dotfill &  8.49$\pm0.046$& 7\\
               ~~~Li $6708\,$\AA\, EW (\AA) \dotfill & 0.212$\pm$0.034$^\star$ & 7\\
               ~~~H$\alpha$ EW (\AA) \dotfill & 0.797$^\star$ & 7\\
\enddata 
\tablenotetext{}{$^\star$\textit{The observed Li is in absorption and H$\alpha$ is in emission}.\\ $^1$\cite{Stassun:2019}; $^2$\cite{2MASS}; $^3$\cite{APASS}; $^4$\cite{GaiaDR3}; $^5$\cite{UCAC}; $^6$\cite{WISE}; $^7$This work; $^{8}$Interpolated from \cite{Claret2017}.
}
\end{deluxetable}

\subsection{Spectroscopic Characterization}

We utilized the publicly available python package \texttt{iSpec} \citep[][]{ispec} to derive spectroscopic stellar parameters from the Veloce spectra for \thisstar\ (see Section~\ref{sec:spec}). We made use of the \texttt{spectrum} synthesis code implemented in \texttt{iSpec}. We fit for [m/H], $T_\mathrm{eff}$, $\log{g}$, and $v\sin{i_\star}$, and used the built-in relations to estimate $v_\mathrm{macro}$ and $v_\mathrm{micro}$. The derived parameters are presented in Table\,\ref{tab:star}. 

The derived parameters ($T_\mathrm{eff} = 3998\pm95$ K, $\log{g} = 4.19\pm0.37$) are in agreement with the empirically derived values from the TIC ($T_\mathrm{eff} = 4000\pm130$ K, $\log{g} = 4.29\pm0.13$). Further, the metallicity derived here (m/H = $-0.08\pm0.11$) is consistent with that of Sco-Cen, which has been shown to be consistent with a solar metallicity \citep[e.g.,][]{Barber:2024}. We adopt the best-fit \texttt{iSpec} values and uncertainties as Gaussian priors to inform our global modeling procedure, which jointly models the planet and stellar parameters (see Section~\ref{sec:global_model}). 

\subsection{\thisstar\ and Sco-Cen}

\begin{figure*}
    \centering
    \includegraphics[width=0.85\linewidth]{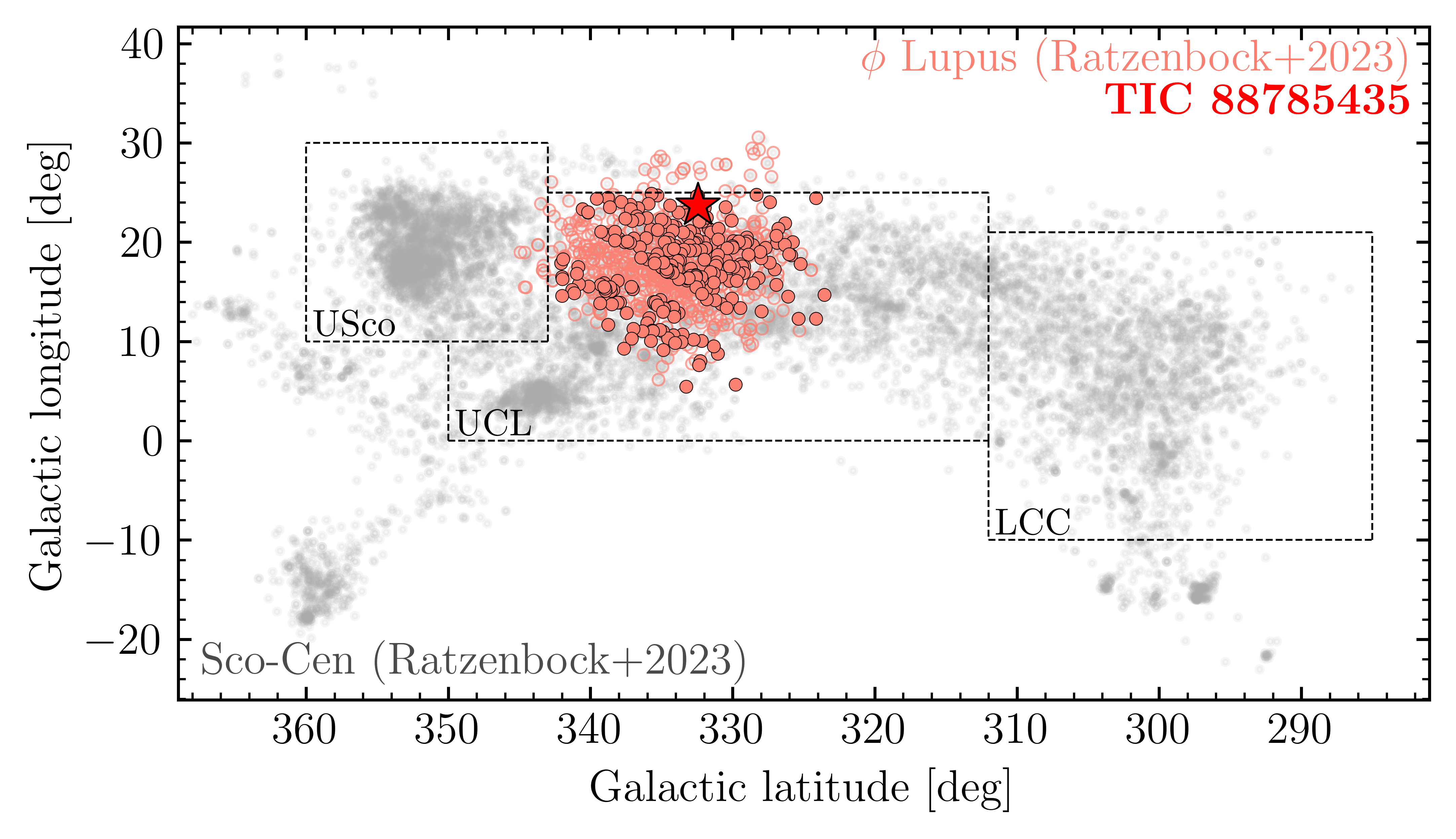}
    \caption{Galactic coordinates ($l,\,b$) of literature Sco-Cen (grey) members \citep[see][and references within]{Luhman:2022_members}{}{}. \thisstar\ (red star) is a member of the classically defined UCL subpopulation. \citet{ScoCen_Sigma:2023} identified \thisstar\ as a high confidence candidate member of the kinematically distinct $\phi$ Lup substructure (salmon circles) primarily within UCL. $\phi$ Lup candidate members outlined in black are rotationally and photometrically validated members of the Sco-Cen population \citep[][]{Rebull:2018, Rebull:2022}{}{}.}
    \label{fig:dist}
\end{figure*}

\thisstar\ has been identified as a candidate member of Sco-Cen \citep{Damiani:2019,ScoCen_Sigma:2023}. Sco-Cen is one of the closest, most well-characterized stellar associations with ongoing star formation. Classically, Sco-Cen is subdivided into three distinct subpopulations based on their galactic coordinates \citep[][]{Zeeuw:1999}{}{}-- Upper Scorpius (USco), Upper Centaurus-Lupus (UCL), and Lower Centaurus-Crux (LCC). 
% While the exact ages of the subpopulations are still an active area of research, the literature agrees the three populations are of different ages, suggesting each comes from a different wave of star formation. 
The literature presents ages for USco at $\mysim8-11\,$Myr \citep[][]{Pecaut:2012}, with UCL and LCC being $\mysim15-20\,$Myr \citep[][]{Mamajek:2002}{}{}. However, the accuracy of this subdivision has long been an area of research, as members within each group have shown evidence of a wide age spread \citep[e.g.,][]{Rizzuto:2015}. Recent works have found that the classical subdivision of Sco-Cen does not fully encompass the complexities and substructures of the association and that Sco-Cen is composed of many more distinct subgroups resulting from many waves of star formation \citep[][]{Goldman:2018,Kerr:2021,Ratzenbock:2023b}. 

\citet{ScoCen_Sigma:2023} performed clustering analysis, using \textit{Gaia}DR3 positions ($X,Y,Z$), tangential velocities ($v_\alpha, v_\delta$), and radial velocities when available with \texttt{SigMA} (Significance Mode Analysis), identifying 37 coeval structures within the population of Sco-Cen. An isochronal age was derived for each substructure, revealing a consistent stream of star formation spanning 3-20 Myr, with four periods of enhanced star formation. \thisstar\ was identified as a candidate member of the $\phi$ Lup (9.9-17.7 Myr) substructure, located within the traditional bounds of UCL (see Figure~\ref{fig:dist}). 

In this work, we attempted to independently validate the membership of \thisstar\ and candidate $\phi$ Lup members. However, due to the limited number of stars that we were able to survey, we were unable to make any statistically significant claims regarding the membership of $\phi$ Lup. In place, we rotationally and spectroscopically confirmed the membership of \thisstar\ to the UCL region and not a contaminating field star. 

% derived isochronal age of 9.9-17.7 Myr for $\phi$ Lup.  

% ages derived  in Here, we attempted to independently characterize the rotational spread, lithium abundance distribution, H$\alpha$, and spectral energy distribution of \thisstar\ and other candidate members of the $\phi$ Lup subpopulation. 

\subsubsection{Stellar Rotations}

Once a low-mass star contracts onto the main sequence, its rotation period can be used to determine its age \citep[][]{Barnes2007, Mamajek2008, bouma2023}{}{}. However, the low-mass stars in UCL are nearly all pre-main-sequence stars and are still undergoing contraction. Therefore, age-rotation relations are unable to measure a precise age \citep[e.g., ][]{Boyle:2023} but the rotation sequence and spread of a coevolved population of stars can still be used to confirm youth. 

\begin{figure}
    % \centering
    \includegraphics[width=\linewidth]{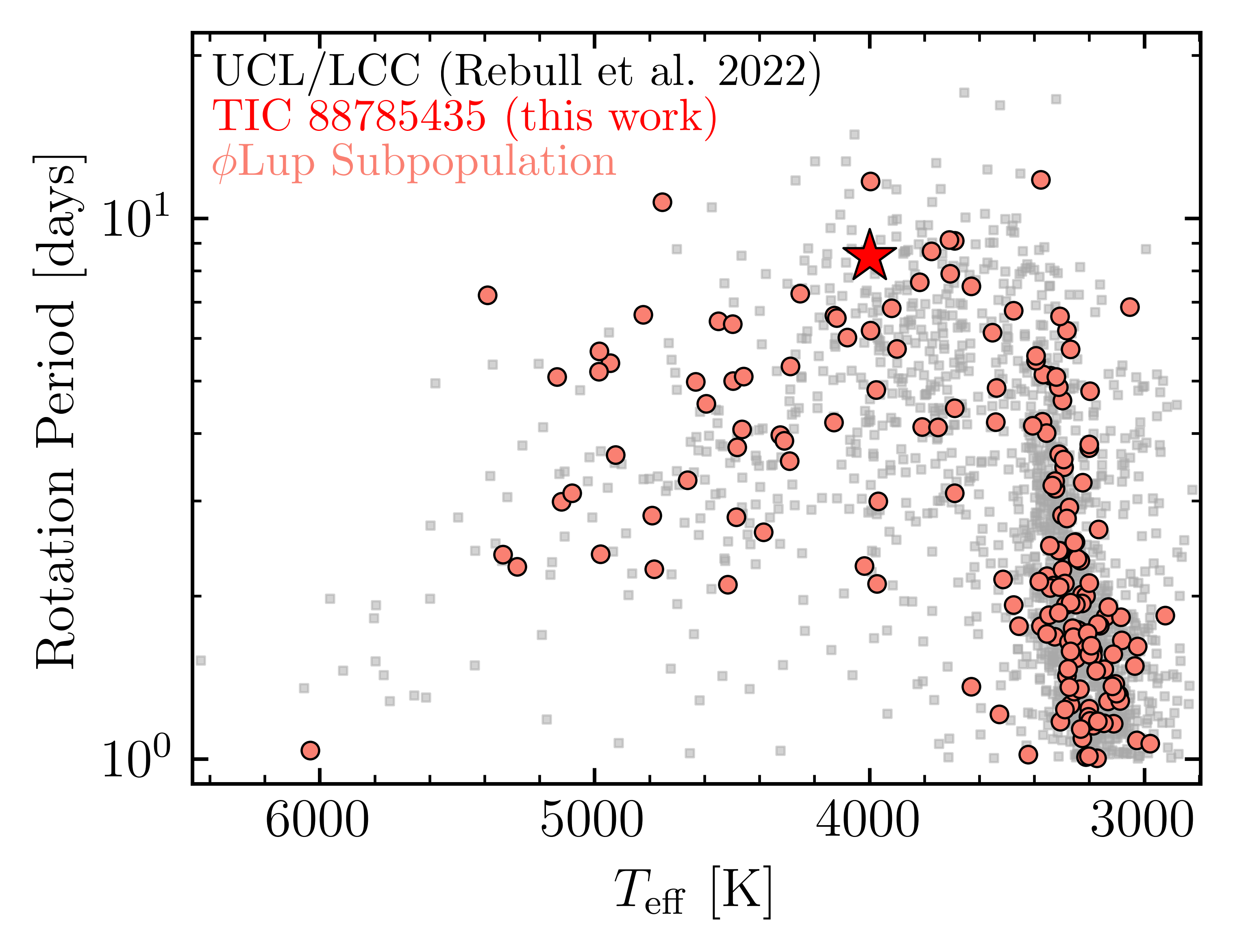}
    \caption{Stellar rotation period as a function of effective temperature for \thisstar\ (red star) from the \tess\ light curves. We plot UCL/LCC members with rotation periods presented in \citet{Rebull:2022} (grey squares). The subset of stars identified as candidate members of the $\phi$ Lup subpopulations are highlighted by the salmon circles.}
    \label{fig:rotp}
\end{figure}

\citet{Rebull:2022} measured the rotation periods of literature members of UCL and LCC with \tess\ light curves (see Figure~\ref{fig:rotp}). Members with \tess\ light curves were classified as either `gold,' `silver,' `bronze,' or `rejected' members. Stars deemed `gold' members were the highest confidence members, and were required to have no source confusion in the \tess\ light curves, $Ks$ and $V$ or $V- Ks$ photometry, and a distance $<300$ pc. \thisstar\ was classified as a `gold' member, with a measured rotation period of 8.4575 days. Ground-based photometry from an NGTS survey studying stellar variability measured a rotation period of 8.44741 days \citep[][]{Briegal:2023}{}{}.

\begin{figure}
    % \centering
    \includegraphics[width=\linewidth]{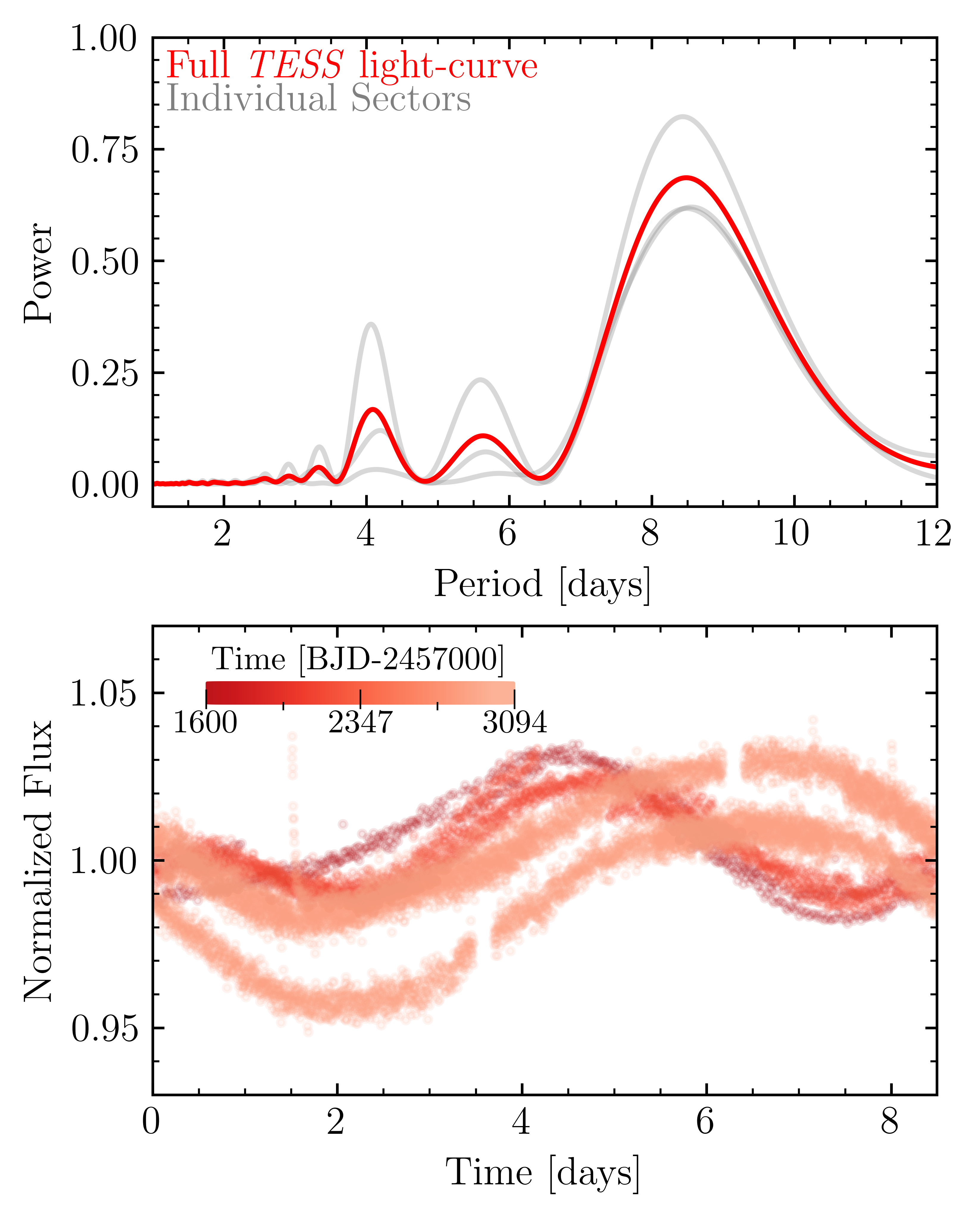}
    \caption{\textit{Upper panel:} Lomb-Scargle periodogram for \thisstar. The Lomb-Scargle for each individual sector is plotted in grey, with the averaged power spectra across all three sectors plotted in red. We measured a rotation period of 8.49 days for \thisstar. \textit{Lower panel:} Phase-folded \tess\ light curve for \thisstar\ at our measured rotation period. }
    \label{fig:ls}
\end{figure}

% We independently measured the rotation distribution of candidate $\phi$ Lup members with \tess. We cross-matched the previously identified candidates of the $\phi$ Lup subpopulation against TICv8.2. We adopted the TIC stellar effective temperatures and filtered for members $T_\mathrm{eff}\leq 6500$\,K. We queried MAST for \tess\ FFI light curves of the selected members. We applied a Lomb-Scargle periodogram to the FFI light curves and searched for rotation periods between 0.5-12 days. Each light curve was visually vetted for aliasing of the true period, binarity, non-rotational variability, and no variability. 
We independently derive the rotation period for \thisstar. We measured a rotation period of $8.49\pm0.046$ days (see Figure~\ref{fig:ls}), implying a rotational velocity of $\mysim5\,\mathrm{km\,s}^{-1}$, which confirms with the derived upper limit on $v\sin{i}$ from the lack of detectable spectral line broadening (see Section~\ref{sec:obs}). We estimate the uncertainty on the derived rotation period by measuring the scatter of the peak in the periodogram across all available sectors. We compared our measured rotations against literature values to ensure consistency. All measured rotation periods agree to within $\mysim0.5$ days. Furthermore, the rotation period of \thisstar\ is consistent with the established distribution of rotation periods for UCL members at similar $T_\mathrm{eff}$ values \citep[][]{Rebull:2022}.

% The variation of the photometric modulation in each sector of the \tess\ light curve (Figure~\ref{fig:tess}) suggests this star has experienced spot evolution since the first \tess\ observations. To verify our rotation measurement, we repeat the analysis only including the first two sectors (the data available for \citet{Rebull:2022}), and for each sector individually. We present our periodograms, and phased \tess\ light curve at the rotation period in Figure~\ref{fig:ls}. 

\subsection{Li and H$\alpha$ equivalent widths}\label{sec:li_halpha}

The spectroscopic features of Sco-Cen members have been well characterized within the literature \citep[e.g., see][and references within]{Luhman:2022_members}. Unfortunately, \thisstar\ was not observed as a part of these previous efforts. We used the spectra of \thisstar\ and the six candidate members of UCL using Veloce (see Section\,\ref{sec:spec}) to investigate their Li and H$\alpha$ features.

\begin{figure*}
    % \centering
    \includegraphics[width=0.48\linewidth]{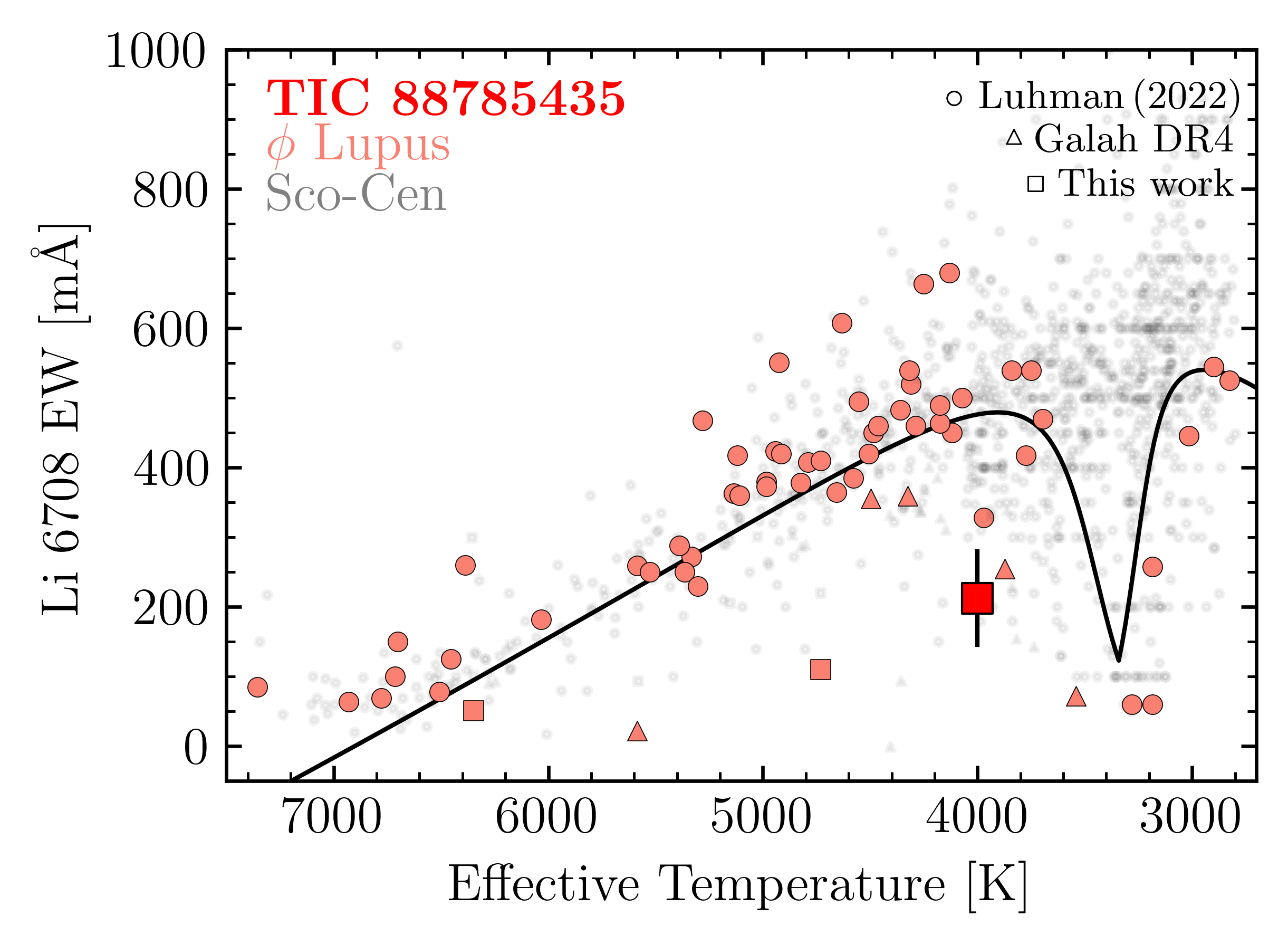}    \includegraphics[width=0.49\linewidth]{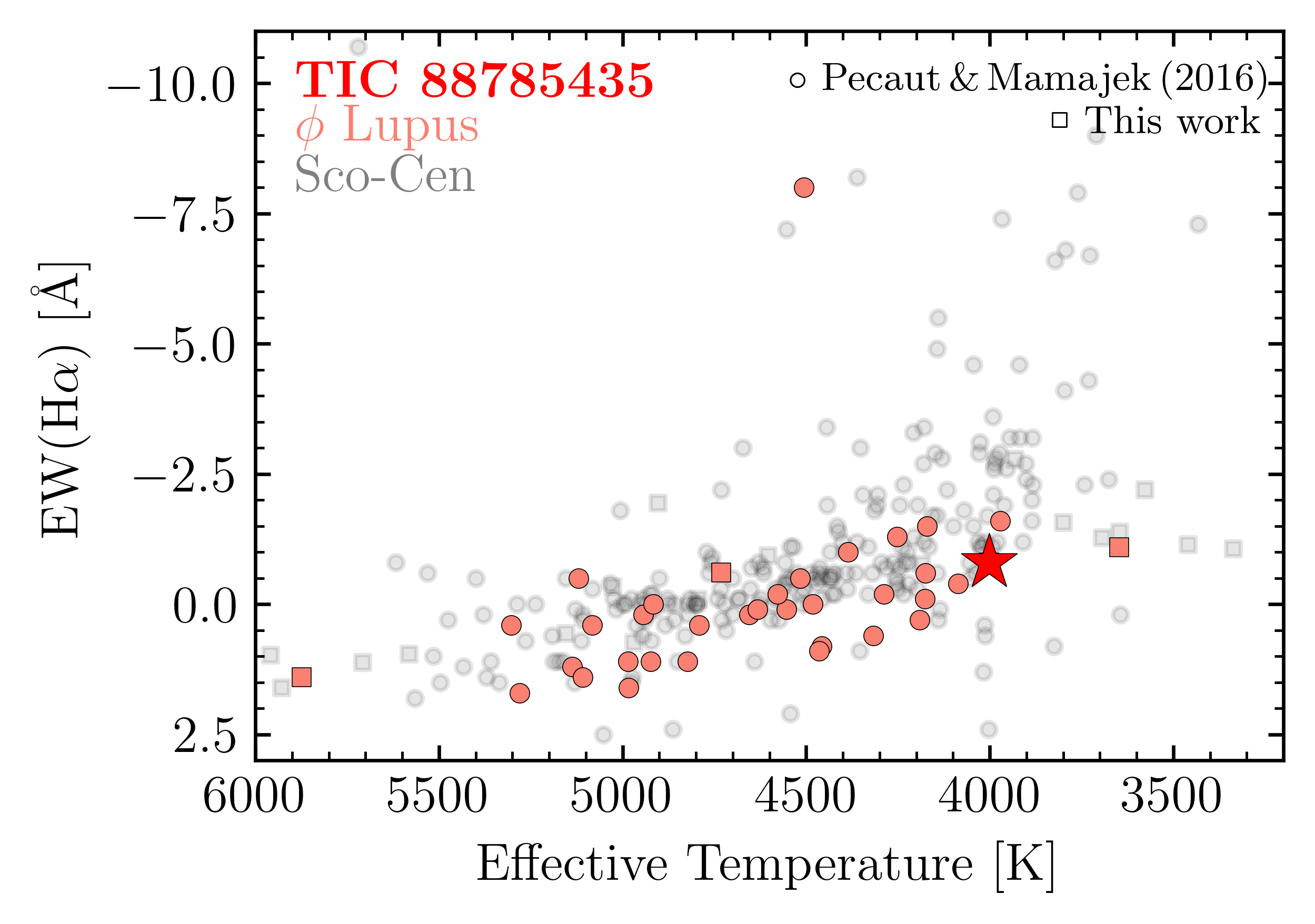}
    \caption{Lithium (left) and H$\alpha$ (right) equivalent widths of TIC 88785435 (red), $\phi$ Lup candidate members (salmon), and literature established Sco-Cen members (grey). We plot the \texttt{EAGLES} \citep[][]{eagles} lithium distribution curve calculated for the age of UCL (16 Myr). Positive values denote the feature was observed in absorption, and negative values indicate emission. \thisstar\ exhibited strong H$\alpha$ emission and Li absorption supporting membership to Sco-Cen. }
    \label{fig:li_ha}
\end{figure*}
% # integrate numerically and multiply by delta lambda 
\begin{table}[ht!]
    \caption{Lithium Equivalent Widths \label{tab:liew}}
    \centering
    \begin{tabular}{lcr}
        \hline\hline
         TIC & Li EW [\AA] & $V$mag\\
        \hline
        \textbf{88785435}\,\, & 0.212$\pm$0.034 &  \,\, 13.259$\pm$0.114\\
         57450390\,\, & 0.24$\pm$0.13 & \,\, 8.62 ± 0.030\\
         160197791\,\, & 0.17$\pm$0.014& \,\, 12.354 ± 0.069\\
         148312561\,\, & 0.112$\pm$0.062 & \,\, 10.913 ± 0.035 \\
         93700013\,\, & 0.072$\pm$0.053 & \,\, 9.07 ± 0.030 \\
         49974963\,\, & 0.136$\pm$0.051 & \,\, 13.606 ± 0.080 \\
         205056288\,\, & 0.118$\pm$0.071 & \,\, 10.403 ± 0.007
 \\
         \hline
    \end{tabular}
\end{table}

We derived the lithium equivalent widths (EW) following the methodology outlined in \citep[][]{Zhou2021}. Briefly, we fit the region of the spectra around the 6708\AA{} Lithium doublet and the nearby Fe I line at 6707.43\AA{}, which often contaminates the Li doublet in young stars, using three Gaussian profiles. We modeled the Li doublet using two Gaussians with equal amplitudes, while the amplitude for the Fe I feature was allowed to vary. We assumed the widths of all three Gaussians were equal to the stellar rotational broadening velocity. We adopt the integral of the Li doublet as the equivalent width. The derived Li EW are presented in Tabel\,\ref{tab:liew}. 

We derived the H$\alpha$ EW following \citet{West:2011,Newton:2017}. Unlike the Li doublet, we do not model the shape of the H$\alpha$ feature but rather adopt the integral within a window centered around the 6563\AA{} H$\alpha$ feature as the EW \citep[see Eq. 1][]{Newton:2017}, with the limits of integration set between 6558 and 6568\AA{}. 

The measured EWs for both Li and H$\alpha$ enable an independent confirmation of youth and association membership. Similarly to the rotation, \thisstar\ agrees with the distribution of Li and H$\alpha$ EWs for previously surveyed members of UCL. Our measured Li and H$\alpha$ EW for \thisstar\ are presented in Figure~\ref{fig:li_ha}. 

\subsubsection{\thisstar\ membership to UCL}
We measured rotation periods and obtained spectroscopic follow-up of candidate members of $\phi$ Lup, including \thisstar, and the wider UCL population. We established, via these observations and characterizations as mentioned above, that \thisstar\ and identified $\phi$ Lup candidates are consistent with the wider UCL population. We therefore adopt an age of $16\pm2$ Myr for the star for the rest of this manuscript as per \citet{Mamajek:2002}. We adopt this age as a prior for the global modeling analysis described below in Section~\ref{sec:global_model}.

\section{Global Modeling}\label{sec:global_model}

\begin{deluxetable}{llcr}
\tablewidth{1pc}
\tabletypesize{\scriptsize}
\tablecaption{
        Best-fit values of \thisstarb% \hatstara{}
    \label{tab:planet}
}
\tablehead{
    &
    &
    &
     \\
    \multicolumn{1}{l}{~~~Paramter}          &
    \multicolumn{1}{l}{~~~}          &
    \multicolumn{1}{c}{Value}            &
    \multicolumn{1}{r}{Prior}            
}
\startdata  
            \,\,\, $T_0$ (BJD) \dotfill & \dotfill&$2458609.2369_{-0.0061}^{+0.0055}$ & Uniform\\
            \,\,\, $P_b$ (days) &\dotfill& $10.508843_{-0.000034}^{+0.000037}$ &Uniform \\
             \,\,\, $R_b/R_\star$ &\dotfill & $0.0508_{-0.0017}^{+0.0018}$ & Uniform\\
             \,\,\, $i$ (degrees) &\dotfill & $88.97_{-0.39}^{+0.73}$ & Uniform\\
             \,\,\, $\sqrt{e}\cos{\omega}$ &\dotfill &$-0.003^{+0.136}_{-0.136}$ &Uniform \\
             \,\,\, $\sqrt{e}\sin{\omega}$ &\dotfill &$-0.010_{-0.061}^{+0.069}$ &Uniform \\[0.04cm]
             \hline 
             \,\,\, $R_b$ ($\mathrm{R_\oplus}$) &\dotfill &$5.03_{-0.20}^{+0.21}$ &Derived \\
             \,\,\, $a/R_\star$ &\dotfill &19.89$\pm0.52$ &Derived \\%[0.095cm]
             \,\,\, $a$ (AU) &\dotfill &$0.08432\pm0.00068$&Derived \\%$[0.095cm]
             \,\,\, \textit{e}& \dotfill &$0.131^{+0.060}_{-0.064}$&Derived \\
             \,\,\, $\omega$ (degrees) &\dotfill &$-25_{-113}^{+157}$ &Derived \\
           \,\,\, $T_\mathrm{eq}$ (K)$^*$ &\dotfill &$635\pm16$ &Derived \\
           \,\,\, $T_\mathrm{14}$ (hours) &\dotfill &$4.10\pm0.15$ &Derived \\
           \,\,\, $T_\mathrm{23}$ (hours) &\dotfill &$3.88\pm0.14$ &Derived \\
\enddata 
\tablenotetext{}{$^*$Assuming zero albedo and the planet retains all irradiation.
}
\end{deluxetable}

We performed a global modeling exercise to derive the best-fit parameters for the \thisstar\ system, simultaneously modeling the planet transits and the stellar SED to best incorporate the uncertainties from each observation. We followed the procedure presented in \citet{zhou:2022}, implementing the methodology outlined in \citet[][]{Vach:2024b}{}{} Section 4.

In short, our free parameters, orbital period, $P$, radius ratio, $R_p/R_\star$, orbital inclination, $i$, time of transit center, $t_0$, and eccentricity parameters, $\sqrt{e}\cos{\omega}$ and $\sqrt{e}\sin{\omega}$, describe the planet transits of \thisstarb. We adopted quadratic limb-darkening parameters $(u_1, u_2)$ and imposed Gaussian priors using the values interpolated from \citet{Claret2017} in the bandpasses of our photometric observations (\tess, Sloan \textit{i'}, and Sloan \textit{g'}). We modeled the transit of \thisstarb\ with \texttt{batman} \citep[][]{batman}{}{}, the python implementation of the transit models presented in \citet{Mandel:2002}. All available \tess\ data and LCO-photometry were used in the global modeling procedure. 

For the \tess\ photometry, we modeled the stellar variability present in the \tess\ light curve via a third-order polynomial and then selected a half-day region around each transit to be used in the transit modeling. At each iteration, we detrended the LCO photometry by calculating and subtracting a trend model from the observations. The trend was fitted to the residuals from the tested transit model subtracted from LCO light curves. The trend was modeled using a linear combination of the airmass, pixel positions $X$, $Y$, and full width at half the maximum of the point spread function. 

The parameters of the host star, \thisstar, were described by our stellar free parameters, stellar mass, $M_\star$, effective temperature, $T_\mathrm{eff}$, metallicity, [m/H], age, and parallax. At each iteration, we estimated the stellar magnitudes by interpolating the stellar parameters onto a MESA Isochrones and Stellar Tracks (MIST) isochrone \citep[][]{Dotter:2016}{}{} via \texttt{minimint} \citep{Koposov:2021}. The estimated stellar magnitudes were then compared against the observed magnitudes of \thisstar\ from \textit{Gaia} DR3 \textit{G, BP,} and \textit{RP} \citep[][]{GaiaDR3}, \textit{WISE} \textit{W1, W2,} and \textit{W3} \citep[][]{WISE}, \textit{2MASS} \textit{J, H,} and \textit{Ks} \citep[][]{2MASS}, and \textit{APASS} \textit{B} and \textit{V} bands \citep[][]{APASS}. The stellar parameters are then used to calculate $a/R_\star$ in the transit model at each iteration. 

We used the measured \textit{Gaia}DR3 parallax and associated uncertainties to impose a Gaussian prior on our derived parallax. Further, we imposed a Gaussian prior on the stellar age, $16\pm2$ Myr, adopted from \citet{Mamajek:2002}. We made use of our spectroscopic characterization (see Section~\ref{sec:spec}) to impose Gaussian priors on the metallicity, and to ensure the derived effective temperature and $\log{g}$ at each iteration are consistent with the spectroscopic observations. The best-fit values and their respective posteriors were then explored through the Markov Chain Monte Carlo implemented in \texttt{emcee} \citep[][]{emcee}. Our best-fit values are presented in Table~\ref{tab:planet}.

\section{Investigating possible false-positive scenarios}\label{sec:fp}

Both astrophysical and instrumental signals can mimic planetary transits in the \tess\ light curves. Here, we explored various false-positive scenarios and derived false-positive probability for \thisstarb.  

Due to the large pixel size of \tess, signals identified as planet candidates may result from contamination of the light curve from nearby stars. Our LCO transit observations confirmed the transit is on target with an uncontaminated 3.9'' aperture and cleared the field of nearby eclipsing binaries $<2.5.'$ There are no \textit{Gaia} resolved stars within 3.9'' of \thisstar, with the nearest neighbor located 7.63'' away (\textit{G}mag = 19.38). 

The LCO detections ruled out instrumental or systematic effects inducing the signal as the transit was detected by multiple instruments. Further, the LCO transits have equal depth at $g'$ and $r'$ bands, confirming that the occulting body is not significantly luminous.

Unresolved background stars can be the source of detected transit signals \citep[][]{Seager:2003}{}{}. We determined the brightest possible magnitude that an unresolved background star would need to be in order to induce the observed transit signal following \citet{Vanderburg:2019} Equation 4. An unresolved star would have to be within $\Delta m \lesssim 2.94$ to be able to be the source of the observed transit signal. However, our high-resolution imaging detected no secondary sources, achieving a contrast ratio of $\Delta m>5$ at 0.1'' which corresponds to a projected separation of 12.2 au (see Section~\ref{sec:hri}).
 % Our high-resolution imaging detected no secondary sources, achieving a contrast ratio of $\Delta m>5$ at 0.1'' (see Section~\ref{sec:hri}). However, unresolved background stars can be the source of detected transit signals \citep[][]{Seager:2003}{}{}. We determined the brightest possible magnitude that an unresolved background star would need to be in order to induce the observed transit signal following \citet{Vanderburg:2019} Equation 4. An unresolved star would have to be within $\Delta m \lesssim 2.94$ to be able to be the source of the observed transit signal.

To estimate the probability of any remaining false-positive scenarios, we performed a joint analysis combining \texttt{MOLUSC} \citep[][]{Wood:2021}{}{} and \texttt{TRICERATOPS} \citep[][]{Giacalone:2021}{}{}. We simulated 100,000 possible binary scenarios with \texttt{MOLUSC}. Our high-resolution imaging and \textit{Gaia} renormalized unit weight errors (RUWE) and photometry were used in the \texttt{MOLUSC} simulations. In turn, the \texttt{MOLUSC} simulations, paired with our follow-up observations, were used to inform our false-positive probability calculation with \texttt{TRICERATOPS},
eliminating physically unrealistic false-postive scenarios. We performed 50 \texttt{TRICERATOPS} calculations to derive a false-positive probability of $<1.40\times10^{-4}$ at the $5\sigma$ level. We, therefore, present \thisstarb\ as a validated planet. 

\section{Preliminary Planet Survey of Sco-Cen}\label{sec:survey}

Since the discovery of the K2-33 b \citep[][]{David:2016, Mann2016}, the first transiting planet $<20$ Myr, the planet population within Sco-Cen has helped to illuminate the observational differences between the newborn planet population and the \textit{Kepler} field \citep[][]{Rizzuto:2017}. %aiding to bridge the gap between observations and our current understanding of planet formation and evolution. 
\textit{Gaia} DR3 spatial positions and proper motions have unveiled the complex and diverse substructures and star formation histories of Sco-Cen \citep[e.g.,][]{Ratzenbock:2023b}, and provided updated membership lists, with an estimated field star contamination rate of $\mysim6\%$ \citep[][]{ScoCen_Sigma:2023}. The increasingly complete membership lists resulting from clustering analyses, paired with the near full-sky coverage of \tess, enabled initial occurrence rate calculations on a per-cluster basis. To date, \tess\ has observed nearly 70\% of the region in which Sco Cen occupies on the sky. Here, we present a preliminary survey of short-period ($<20$ days) planets in Sco-Cen with \tess\ and provide an initial forward modeling comparison against mature-aged demographics following the methodology outlined in \citet{Vach:2024}. 

We made use of stable Sco-Cen members identified in \citet[][]{ScoCen_Sigma:2023} as our parent stellar population. We cross-matched stable Sco-Cen members with TICv8 \citep[][]{Stassun:2019}, and adopted TICv8 stellar parameters for all subsequent analyses. We selected for $T_\mathrm{eff} < 7000$ K and $M_\star< 2 \mathrm{M_\odot}$, and attempted to filter for binarity, removing stars with Gaia DR3 RUWE $>1.4$ following \citet{Kervella:2022}. This yields 5713 members with \tess\ QLP FFI observations (from here on this subset is referred to as QLP). The QLP data were made available from the Mikulski Archive for Space Telescopes (MAST) as of 27 June 2024. 

As young stars present unique challenges when searching for transiting planets due to their heightened photometric variability, we first detrended for stellar activity in the \tess\ light curves and then performed a planet search. Our planet search and vetting routine is outlined in short above (see Section~\ref{sec:obs}), and in-depth in \citet{Vach:2024} Section 4. Unlike in \citet{Vach:2024}, here we scale the parameters of the spline to optimize planet recoverability on a star-by-star basis. Optimized detrending and outlier rejection break spaces are selected based on our injection and recovery simulations as per \citet{Vach:2024} Section 5.

We identified 6 TCEs that passed all our vetting procedures. All TCEs were then searched for multiplicity. We masked the initial signal that triggered the TCE in the raw light curve and then repeated our planet search. An additional TCE was identified in the HIP 67522 \tess\ data, which passed our vetting procedures. Our recovered planet population is presented in Table~\ref{tab:planets}, including 4 confirmed planets, 1 TOI candidate, and 2 planet candidates not previously identified as TOIs. As this is a preliminary survey, no follow-up has been completed in an attempt to rule out false positives. We, therefore, followed the methodology outlined in Section~\ref{sec:fp} to estimate the false-postive probability of each candidate. We present our false-positive probabilities in Table~\ref{tab:planets}. We note no newly identified candidate was able to be statistically validated using \tess\ data alone-- i.e. a false-positive probability $< 0.015$ as per \citet{Giacalone:2021}.  We expand upon each non-TOI candidate in Appendix~\ref{sec:candidates}. 

% \begin{table*}[ht!]
%     \centering
%     \caption{Planets and candidates identified in our search}
%     \label{tab:tois}    
% \begin{tabular}{llccccccc}
% \toprule \toprule
%   TIC &             Name & $R_P$ ($R_\oplus$) & $P$ (days) &    $T_0$ (BTJD) &  $T_\mathrm{eff}$ (K) &  Age (Myr) & Subgroup (\texttt{SigMA}) & False-positive Rate\\
% \midrule \midrule
% \multicolumn{9}{l}{\textbf{Confirmed and verified planets}}\\ 
% 88785435 & TIC 88785435 b$^a$ & 4.88$\pm$0.28 & 10.51 & 2457000.0000 & 4000 & 16 & $\phi$ Lup (15) & \nodata \\
%  166527623 & HIP 67522 b$^a$ & 4.88$\pm$0.28 & 10.51 & 2457000.0000 & 4000 & 16 & $\nu$ Cen (20) & \nodata \\
% 88785435 & TIC 88785435 b$^a$ & 4.88$\pm$0.28 & 10.51 & 2457000.0000 & 4000 & 16 & $\nu$ Cen (20) & \nodata \\

% \hline
% \multicolumn{8}{l}{\textbf{Planet candidates}}\\ 

% 88785435 & TIC 88785435.01$^i$ & 4.88$\pm$0.28 & 10.51 & QLP & 4000 & 16 & UCL & $0.0947\pm0.0037$ \\

% \hline
% \end{tabular}

% % $(a)$ This work.
% % $(\alpha)$ No signatures of youth in spectroscopic follow-up; ($\beta$) Not in our survey parameters; $(\gamma)$ Low SNR, ambiguous signal-- not a TOI as of February 2023; 
% % $(\delta)$ Ambiguous Planet Candidate; $(\epsilon)$ Nearby Eclipsing Binary SG2; $(\zeta)$ Nearby Planet Candidate; $(\eta)$ Eclipsing Binary SG2; .\\
% % $^*$The pipeline column identifies which of our surveys recovered the signal.
% \end{table*}

\begin{deluxetable*}{rlccccccr}

\tablewidth{0pc}
\tabletypesize{\scriptsize}
\tablecaption{
        Recovered Planets of Sco-Cen% \hatstara{}
    \label{tab:planets}
}
\tablehead{
    &
    &
    &
    & \\
    \multicolumn{1}{l}{TIC}          &
    \multicolumn{1}{l}{Planet Name}            &
    \multicolumn{1}{c}{$R_P$ ($R_\oplus$)}            &
    \multicolumn{1}{c}{$P$ (days)}            &
    \multicolumn{1}{c}{$T_\mathrm{eff}$ (K)} &
    \multicolumn{1}{c}{Age (Myr)} &
    \multicolumn{1}{c}{Subgroup (\texttt{SigMA})} &
    \multicolumn{1}{r}{False-positive Rate} &
}
\startdata  
\multicolumn{9}{l}{\textbf{Confirmed and verified planets}}\\ 
88785435 & TIC 88785435 b$^a$ & 5.03$\pm$0.21 & 10.51 &  4000 & 15.96±1.60$^{a}$& $\phi$ Lup (15) & \nodata \\
 166527623 & HIP 67522 b$^{b,c}$ & 9.99$\pm$0.24 & 6.96 & 5675 & $17\pm2^{b}$ & $\nu$ Cen (20) & \nodata \\
 166527623 & HIP 67522 c$^c$ & 7.94$\pm$0.36 & 14.33 &  5675 & $17\pm2^{b}$ & $\nu$ Cen (20) & \nodata \\
455000299 & TIC 455000299 b$^d$ & 4.72$\pm$0.30 & 18.71 &  4400  & 6.7-11.8$^f$ & Musca-foreground (23) & \nodata \\
\hline
\multicolumn{9}{l}{\textbf{Planet candidates}}\\ 
% 359357695 (CHECK APOGEE FOR THE NORTH) & TOI 1881.01$^i$ & $\pm$ &  &  &  &  6.7-11.8 & Musca-foreground (23) &  \\
210904767 & TOI 7038.01$^e$ & 3.41$\pm$0.32 & 8.59 & 6920  & 8.8 - 15.5 & Scorpio-Body (8) & 0.0121$\pm$0.0025\\
273586149 & TIC 273586149.01$^a$ & 6.20$\pm$0.48 & 5.17 &  3970  & 8.8-16.1 & $\sigma$ Cen (21) & 0.38±0.26\\
% 75379256 (ONLY TWO TRANSITS) & TIC 75379256.01$^i$ & $\pm$ &  &  &  & 9.4-17.9 & $\phi$ Lup (15) &\\
% 404226094 & TIC 404226094.01$^i$ & $\pm$ &  &  &  & 8.8-16.1 & $\sigma$ Cen (21) &\\
89071445 & TIC 89071445.01$^a$ & 5.42 $\pm$0.40 &   3.27 & 4110  & 9.4-17.9 & $\phi$ Lup (15) & 0.0417±0.0015\\
\enddata 
\tablenotetext{}{$^a$This work; $^b$\citet{Rizzuto:2020}; $^c$\citet{Barber:2024}; $^d$Ringham et al. \textit{in preparation}; $^e$ExoFOP; $^f$\citet{Ratzenbock:2023b}.
}
\end{deluxetable*}

% 455000299 & TIC 455000299.01$^i$ & $\pm$ &  &  &  & 16 & UCL &\\

We note the exclusion of K2-33 \citep[][]{David:2016,Mann2017}, TOI-1227 b \citep[][]{Mann:2022} and HD 109833 \citep[][]{Wood:2023} planetary systems from recovered planet population. K2-33, an established literature member of Upper Sco, has yet to be observed by \tess\ (observations predicted for Year 7, Sector 91) and is therefore not a part of our preliminary Sco-Cen planet search. TOI-1227 b is a sub-Jovian-sized planet orbiting an established literature member of Sco-Cen. Its orbital period is outside our parameters, 27.4 days, and therefore not included in our search. HD 109833 was previously identified to be associated with LCC. However, \citet{ScoCen_Sigma:2023} did not identify HD 109833 as a member of Sco-Cen, therefore, it was not included in our parent stellar population. TOI-1881.01 was in our parent sample but was ruled a spectroscopic false-positive and is therefore not included in Table~\ref{tab:planets}. 

This preliminary survey identified multiple Neptune and Jovian-sized planets with short orbital periods, which have been found to be rare within the Kepler population \citep[$<2\%$][]{Kunimoto2020}. Previous theoretical and population studies have suggested that newborn planets with larger radii are common at ages $<50$ Myr, as they are small-planet progenitors still inflated from the process of formation \citep[e.g.,][]{owen:2013,Owen:2016, Rogers:2021, Rogers:2023, Fernandes:2022, Vach:2024, Karalis:2025}. The young sub-Jovian sized planets that have thus far been observed with JWST and HST, HIP 67522 b \citep[located in Sco-Cen,][]{Thao:2024} and V1298 Tau b and c \citep[30 Myr,][Barat et al. \textit{in prep.}]{Barat:2024a, Barat:2024b}, have been found to have small core masses, more consistent with that of a super-Earth or a mini-Neptune rather than a giant planet, engulfed in an extended atmosphere. While these observations are consistent with a gas-rich planet formation scenario, where planets are born inflated, engulfed in a light, extended atmospheric envelope that then both contracts and is stripped away due to cooling and interactions with the young star \citep[][]{Rogers:2021,Rogers:2023}, a statistically significant sample of young planets (<50 Myr) with mass constraints is required to distinguish between the small planet progenitor and the young, close-in massive planet scenarios. As the population surveyed here, including \thisstarb\ and HIP 67522 b, is younger than 20 Myr, it provides an ideal laboratory to test the nature of newborn short-period planets through future atmospheric characterization and dynamical mass characterization.

\section{Summary and Conclusions}\label{sec:discussion}

In this paper, we presented the discovery and characterization of \thisstarb, a newborn, transiting super-Neptune located in the Sco-Cen OB association. \thisstar\ was originally identified as a planet host in \citep[][]{Vach:2024}. We investigated the Li and H$\alpha$ equivalent widths of \thisstar\ to further support its membership. We detected the lithium absorption feature and H$\alpha$ in emission, supporting its youth and membership to UCL. Our global model, including \tess\ and ground-based observations, finds \thisstarb\ has a radius of \radb, and an orbital period of \shortperb. With an equilibrium temperature of $635$ K, \thisstarb\ is one of the coolest, newborn ($<30$ Myr) transiting exoplanets amenable to atmospheric characterization.  %Our isochronal fit to the spectral energy distribution of \thisstar\ measured an age of 15.96$\pm1.60$ Myr.   

\thisstarb\ joins the growing number of planets identified orbiting Sco-Cen members. Initial atmospheric characterization and demographic studies of the youngest transiting exoplanets have illuminated a population of inflated young planets. Our preliminary survey of the Sco-Cen planet population, including \thisstarb, further supports this conclusion, finding an excess of larger planets compared to the Kepler demographics of the mature planet population. However, dynamical mass constraints and/or atmospheric characterization of these newborn planets with JWST and ground-based facilities will enable us to directly test whether these planets are small planet progenitors or the youngest glimpses at short-period, Jovian-massed planets.

%%%%%%%%%%%%%%%%%%%%%%%%%%%%%%%%%%%%%%%%%%%%%%%%%%
\section*{Data Availability}

All \tess\ data products used in this paper are publicly available through the Mikulski Archive for Space Telescopes (MAST). Ground-based observations used in this paper are available via ExoFOP-TESS.
 
% The inclusion of a Data Availability Statement is a requirement for articles published in MNRAS. Data Availability Statements provide a standardised format for readers to understand the availability of data underlying the research results described in the article. The statement may refer to original data generated in the course of the study or to third-party data analysed in the article. The statement should describe and provide means of access, where possible, by linking to the data or providing the required accession numbers for the relevant databases or DOIs.

\section*{Acknowledgements}
We respectfully acknowledge the traditional custodians of the lands on which we conducted this research and throughout Australia. We recognize their continued cultural and spiritual connection to the land, waterways, cosmos, and community. We pay our deepest respects to all Elders, past, present, and emerging, and the people of the Giabal, Jarowair, and Kambuwal nations, upon whose lands this research was conducted.

We thank the anonymous referee for their helpful comments and constructive feedback, which have greatly improved the quality of this manuscript. GZ thanks the support of the ARC DECRA program DE210101893 and ARC Future program FT230100517.
CH thanks the support of the ARC DECRA program DE200101840.

Funding for the \tess\ mission is provided by NASA's Science Mission directorate. This research has made use of the Exoplanet Follow-up Observation Program (EXOFOP) website, which is operated by the California Institute of Technology, under contract with the National Aeronautics and Space Administration under the Exoplanet Exploration Program. This paper includes data collected by the \tess\ mission, which are publicly available from the Mikulski Archive for Space Telescopes (MAST).

This work makes use of observations from the Las Cumbres Observatory global telescope (LCOGT) network. Part of the LCOGT telescope time was granted by NOIRLab through the Mid-Scale Innovations Program (MSIP). MSIP is funded by the NSF.

\facilities{\tess. LCOGT.}

\software{\texttt{astropy} \citep{astropy, Astropy:2018, Astropy:2022}\texttt{AstroImageJ} \citep{Collins:2017}, \texttt{batman} \citep{batman}, \texttt{emcee} \citep{emcee}, \texttt{Matplotlib} \citep{matplotlib}, \texttt{numpy} \citep{numpy}, \texttt{scipy} \citep{scipy}, \texttt{TAPIR} \citep{Jensen:2013}, and \texttt{TRICERATOPS} \citep{triceratopscode}.}

\bibliography{refs}{}
\bibliographystyle{aasjournal}

\appendix
\section{Sco-Cen Planet Candidates}\label{sec:candidates}
For each candidate identified in this work, we performed a global model, simultaneously modeling the stellar and planetary parameters with \texttt{emcee} (see Section~\ref{sec:global_model}). Each candidate has been submitted to ExoFOP as a CTIO, along with the best-fit planet parameters. Here, we present our vetting summaries for each candidate and best-fit parameters. 

As a part of our vetting routine, we attempt to ensure the detected signal is consistent with an on-target, planetary transit. Due to the observing strategy of TESS, various astronomical false positives can mimic transit signals consistent with planetary signals, such as nearby eclipsing binaries (NEBs). Oftentimes, bright NEBs are found in nearby pixels, which can then bleed into the neighboring pixels. We attempt to rule out NEB scenarios from our analysis and ensure the observed signal is consistent with being on target by performing a visual per-pixel light curve analysis for each sector of observation. To do so, we extract the light curve of each pixel surrounding the target, detrend for variability in the light curves using a third order-polynomial, and phase fold using the TCE transit epoch and period. We then plot the phase-folded light curve of each pixel centered around the pixel hosting our target star. These plots are presented in our vetting plots (right most panel).

\subsection{TIC 89071445.01}
\begin{figure}
    \centering
    \includegraphics[width=\linewidth]{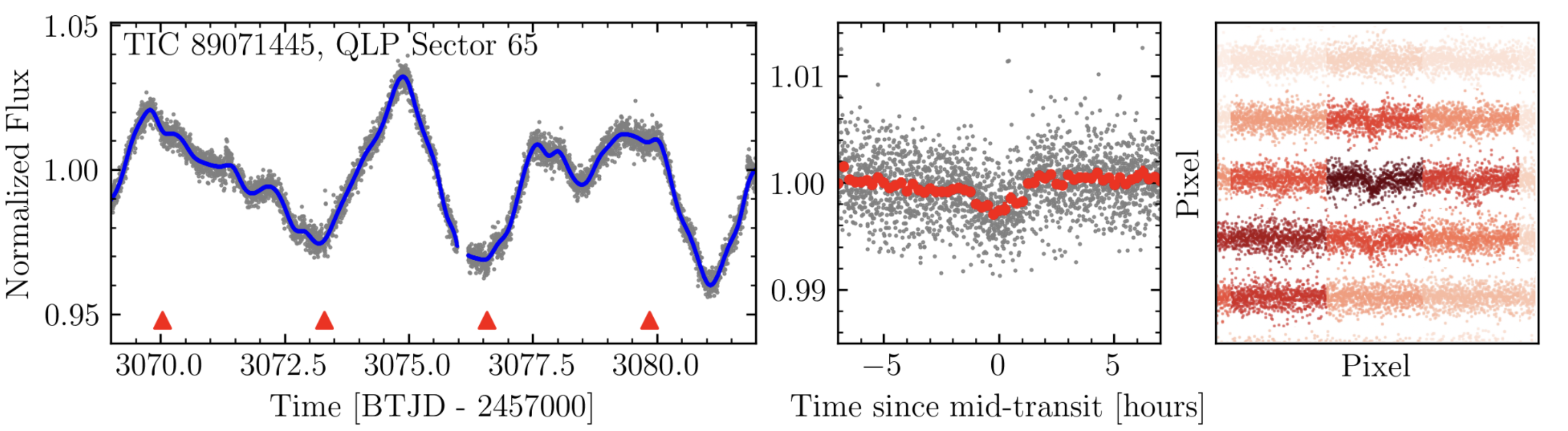}
    \caption{Vetting plot for the Sector 65 observations of TIC 89071455. The \tess\ light curve is presented in the left panel, with the transit times marked by the red triangles. The stellar variability is modeled via a spline (blue). The phase-folded light curve using the best-fit period and transit time is shown in the middle panel. Our per-pixel analysis (right panel) shows the transits are consistent with being on target.}
    \label{fig:candidate1}
\end{figure}
TIC 89071445 ($M_\star = 0.64\,M_\odot$, $T_\mathrm{eff} = 4110 $ K, $T$mag = 11.2) is a candidate member of $\phi$ Lup. \citet{Ratzenbock:2023b} estimates an age of 9.4-17.9 Myr. TIC 89071445 was observed across Sectors 38 and 65 in the TESS FFIs. A 3.27 day signal was identified in our planet search with a signal-to-pink noise of 12.6 (see Figure~\ref{fig:candidate1}). As this star is located within the same subpopulation as \thisstar, we used our best-fit age and associated uncertainties to impose Gaussian priors in the global modeling. We found TIC 89071445.01 is a transiting super-Neptune candidate ($R_p = 5.42\pm 0.40\, R_\oplus$, $P = 3.271945\pm0.000012$ days, $T_0 = 2459337.1144\pm 0.0010$ BJD).

\subsection{TIC 273586149.01}
\begin{figure}
    \centering
    \includegraphics[width=\linewidth]{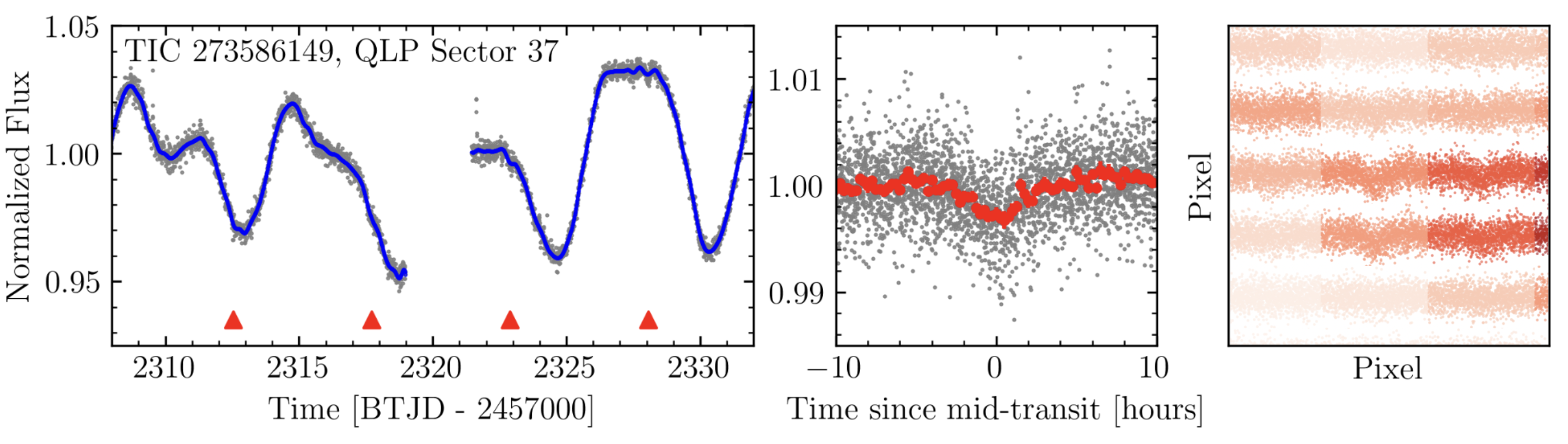}
    \caption{Vetting plot for the Sector 37 observations of TIC 273586149 (same as Figure~\ref{fig:candidate1}).}
    \label{fig:tic273}
\end{figure}
TIC 273586149 ($M_\star = 0.62\,M_\odot$, $T_\mathrm{eff} = 3970 $ K, $T$mag = 10.84) is located in the subgroup $\sigma$ Cen (SigMA 21), with an isochronal age of 8.8-16.1 Myr. TIC 273586149 was observed across Sectors 11, 37, 38, and 64 in the TESS FFIs. Our planet search identified a 5.17 day periodic signal with a signal-to-pink-noise of 15.42 (see Figure~\ref{fig:tic273}). We used the isochronal ages derived from \citet{Ratzenbock:2023b} to inform our global model. Our global modeling procedure finds TIC 273586149 hosts a transiting super-Neptune ($R_p = 6.20\pm 0.48\, R_\oplus$, $P= 5.173450 \pm0.000091$ days, $T_0 = 2458603.769\pm0.024$ BJD).

\section{Comparison of \tess\ data products}\label{sec:candidates}
\begin{figure}[ht!]
    \centering
    \includegraphics[width=0.48\linewidth]{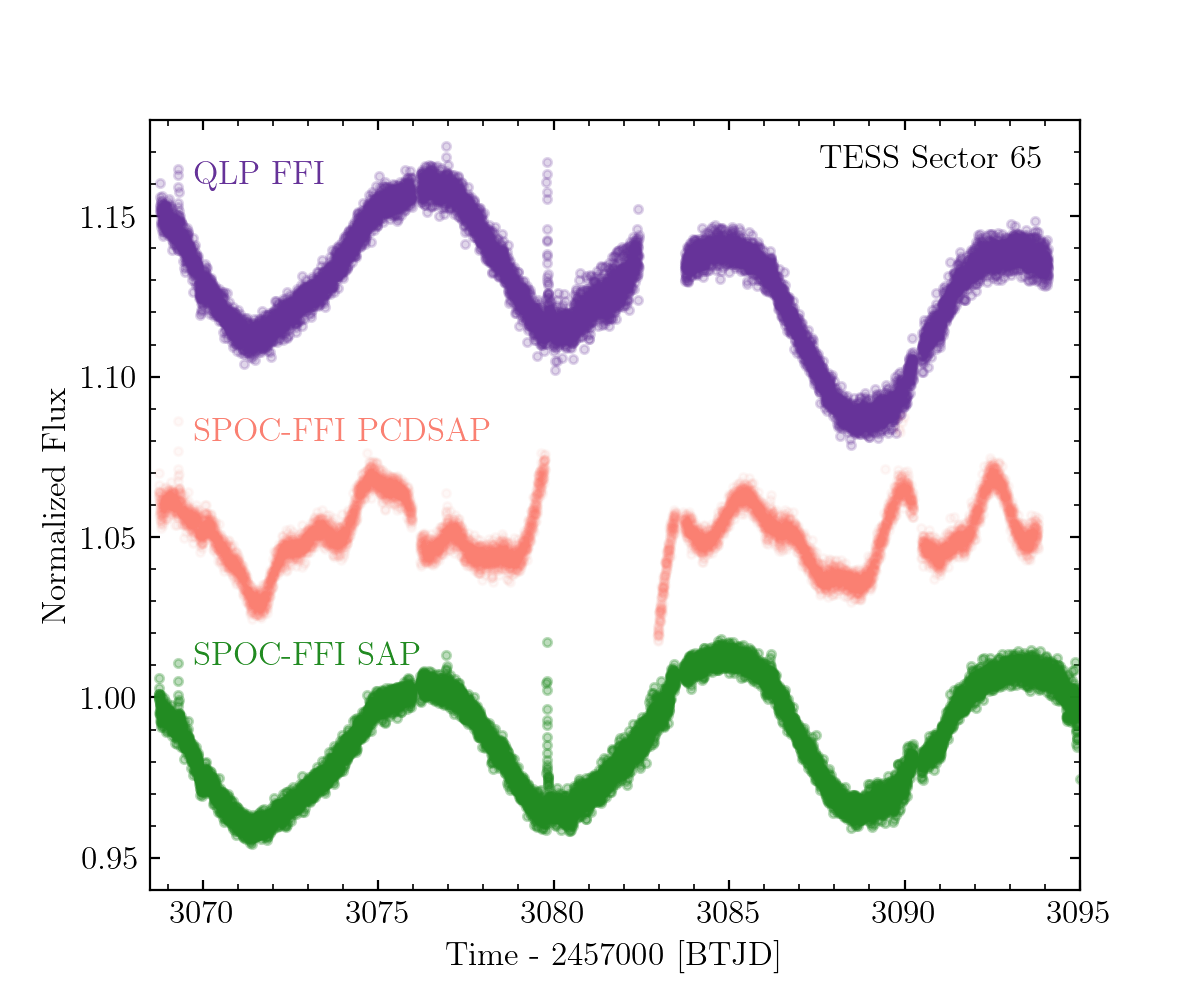}
        \includegraphics[width=0.48\linewidth]{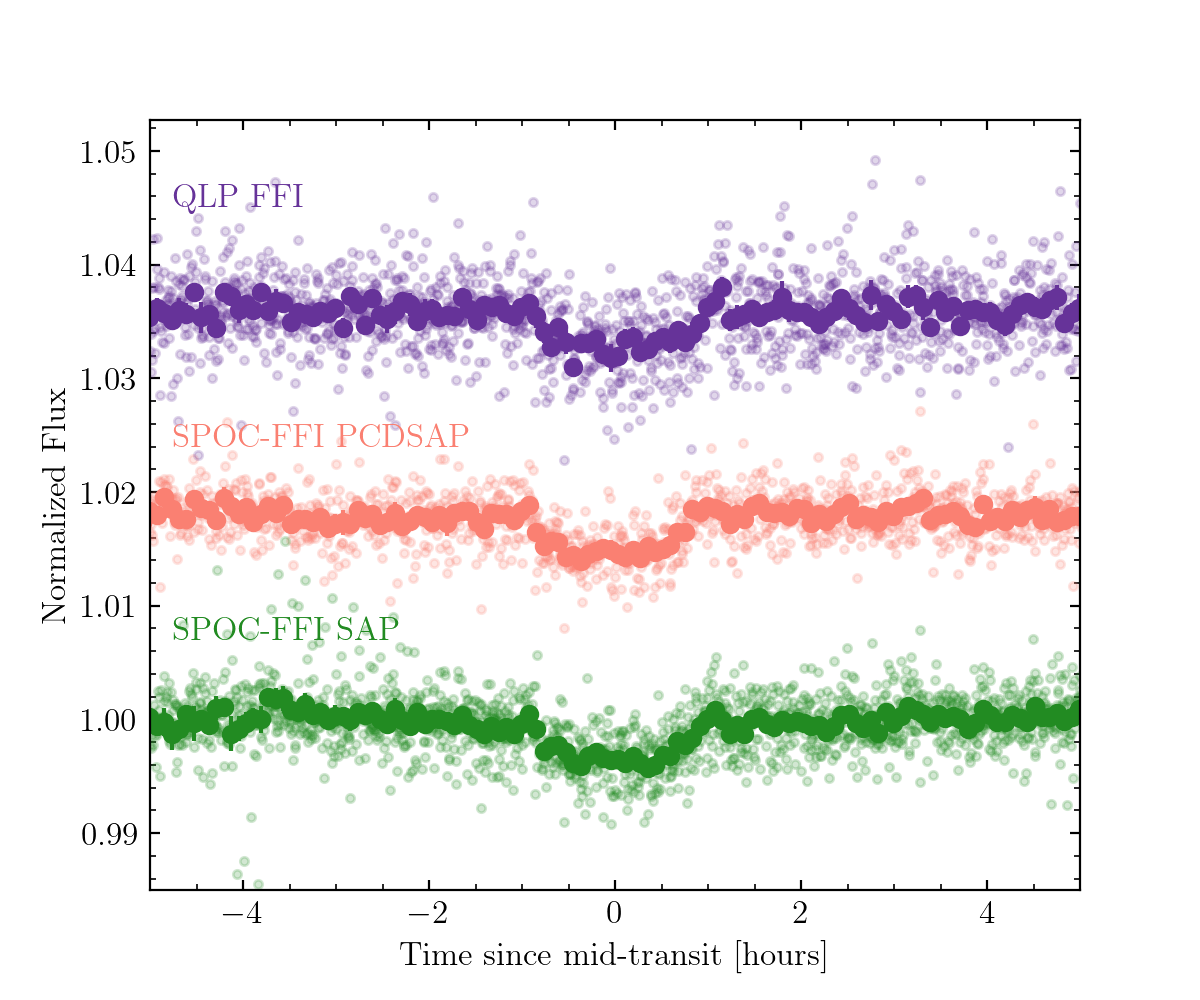}

    \caption{\textit{Left panel}: \tess\ Sector 65 QLP (purple), SPOC FFI PDCSAP (salmon), and SPOC FFI SAP (green) light curves of \thisstar. \textit{Right panel}: Phase folded \tess\ QLP (purple), SPOC FFI PDCSAP (salmon), and SPOC FFI SAP (green) light curves of \thisstarb.}
    \label{fig:tess_comp}
\end{figure}

In this work, we made use of the QLP Full Frame Images of \thisstar\ for our analysis. 
Now in the extended mission, the TESS Science Processing Operations Center \citep[SPOC;][]{jenkins2010tps, SPOC,jenkins2020tps} has also produced FFI light curves \citep[][]{SPOC_FFI}. These data are also publicly available via the Barbara A. Mikulski Archive for Space Telescopes (MAST). For reference, we present a comparison between the Sector~65 light curves from SPOC SAP, SPOC PDCSAP, and QLP, and the resulting phase-folded light curves to illustrate the consistent transit depth across the data products (Figure~\ref{fig:tess_comp}). 

\end{document}